\documentclass[aps,preprint,onecolumn,secnumarabic,nobalancelastpage,amsmath,amssymb,
nofootinbib]{revtex4}

\RequirePackage{fix-cm}
\usepackage{enumerate}
\usepackage[scriptsize,nooneline,hang]{caption}
\usepackage[hang,nooneline,scriptsize]{subfigure}

\usepackage{graphics}      
\usepackage{graphicx}      
\usepackage{url}           
\usepackage{dcolumn}
\usepackage{hyperref} 
\usepackage{bm}            
\usepackage{mathrsfs}
\usepackage{epstopdf}
\usepackage{hyperref}
\usepackage{color}
\usepackage{amsthm,amssymb,amsmath}

\begin{document}

reprint{APS/123-QED}

\title{Thermodynamic geometry of black holes in $f(R)$ gravity}

\author{Saheb Soroushfar}
\author{Reza Saffari}%
\email{rsk@guilan.ac.ir}
\author{Negin Kamvar}
 \affiliation{Department of Physics, University of Guilan, 41335-1914, Rasht, Iran.}%



\date{\today}

\begin{abstract}
In this paper, we consider three types (static, static charged and
rotating charged) of black holes in $f(R)$ gravity. We study the
thermodynamical behavior, stability conditions and phase transition
of these black holes. It will be shown that, the number and type of
phase transition points are related to different parameters, which
shows the dependency of stability conditions to these parameters.
Also, we extended our study to different thermodynamic geometry
methods (Ruppeiner, Weinhold and GTD). Next, we investigate the
compatibility of curvature scalar of geothermodynamic methods with
phase transition points of the above balck holes. In addition, we
point out the effect of different values of spacetime parameters on stability conditions
of mentioned black holes.
\end{abstract}

\maketitle

\section{INTRODUCTION}

The black hole is one of the most fascinating anticipations of
Einstein's theory of General Relativity, which has been an adsorbent
subject in theoretical physics for many years, and it has unknown
issues yet. One of the most interesting aspects of studying black
holes, is thermodynamics. The studies on black holes as a
thermodynamic system is started with famous work of Hawking and
Bekenstein~\cite{Hawking:1974rv,Bekenstein:1973ur,Hawking:1974sw},
which is followed by other pioneering research of
Padmanabhan~\cite{Kothawala:2008in},\cite{Padmanabhan:2003pk}.
According to the black hole thermodynamics, the thermodynamic
quantities of a black hole such as entropy and temperature are
related to it's geometrical quantities such as horizon area and
surface gravity~\cite{Bekenstein:1973ur},~\cite{Bekenstein:1972tm}.
In recent years, the researches on the thermodynamic properties of
the black holes have got a lot of interesting aspects. One of these
aspects is stability of black holes. Heat capacity of a black hole
must be positive in order to be in thermal
stability~\cite{Myung:2007my}. Studying the heat capacity of a black
hole provides a mechanism to study the phase transitions of the
black holes. There are two types of phase transition; in the first
one, the changes in the sign of the heat capacity denoted as a type
of phase transition, in the other words, roots of the heat capacity
represent phase transition points, so we call these phase transition
type one. Another kind of phase transition obtains from divergencies
of the heat capacity. This kind of phase transition is called the
phase transition type two~\cite{Myung:2007my}. Some works on the
normal thermodynamics of black holes shows that in many cases, one
can not identify the detailed reasons irregularities of mass,
temperature and heat capacity shown by the system. During the last
few decades, many efforts have been made to introduce different
concepts of geometry in to ordinary thermodynamics.
Hermann~\cite{Hermann:1973} defined the implication of thermodynamic
phase space as a differential manifold with a natural contact
structure, in which there exist a special subspace of thermodynamic
equilibrium states. Weinhold introduced an other geometric method in
1975~\cite{Weinhold:1975}, in which a metric is defined in the space
of equilibrium states of thermodynamic systems. Weinhold used the
notion of conformal mapping from the Riemannian space to
thermodynamic space. Weinhold's metric is defined as the Hessian in
the mass representation as follows
 \begin{equation}\label{Weinhold}
 g^{W}_{ij}=\partial_{i}\partial_{j}M(S,N^{r}),
 \end{equation}
where $ M $ is the mass, $ S $ is the entropy and $ N^{r} $ is the
other extensive variables of the system. After that in 1979,
Ruppeiner~\cite{Ruppeiner:1979} defined a new metric which is the
minus signed Hessian in entropy representation and is given by
\begin{equation}\label{Ruppiener}
g^{R}_{ij}=-\partial_{i}\partial_{j}S(M,N^{r}).
\end{equation}

The Ruppeiner's metric is conformaly related to Weinhold's metric as
follows~\cite{Mrugala:1984},~\cite{Salamon:1984}
\begin{equation}
ds^{2}_{R}=\frac{1}{T}ds^{2}_{W},
\end{equation}
where $ T $ is the temperature of the thermodynamic system.

Geometrothermodynamics~(GTD) is the latest attempt in this
way~\cite{Quevedo:2006xk},~\cite{Quevedo:2007mj}.
Quevedo~\cite{Quevedo:2006xk} introduced a general form of the
legender invariant metric. The general form of the metric in $ GTD $
method is as follows
\begin{equation}\label{GTD}
g=(E^{c}\frac{\partial\Phi}{\partial E^{c}})(\eta_{ab}\delta^{bc}\frac{\partial^{2}\Phi}{\partial E^{c}\partial E^{d}}dE^{a}dE^{d}),
\end{equation}
in which
\begin{equation}
\frac{\partial\Phi}{\partial E^{c}}=\delta_{cb}I^{b},
\end{equation}
where $ E^{a} $ and $ I^{b} $ are the extensive and intensive
thermodynamic variables and $ \Phi $ is the thermodynamic potential.

There were some alternative and extended theories on General
Relativity from the beginning it \cite{Weyl:1919fi,Eddington:1923,Brans:1961sx}. Some of new versions of
these theories are trying to justify some observed anomalies in
galactic scales (dark matter) and cosmological scales (dark energy)
which leads to reinforce them, such as, scalar-tensor theories,
brane world cosmology, Lovelock gravity and $ f(R) $ gravity. Many
different aspects, such as, cosmic inflation, cosmic acceleration,
dark matter, correction of the solar system abnormalities, and also
geodesic motion of test particle, have been studied in $ f(R) $
gravity
\cite{Brans:1961sx,Riess:1998cb,Fujii:2003,Brax:2003fv,Lovelock:1971yv,Buchdahl:1983zz,Starobinsky:1980te,Bamba:2008ja,Akbar:2006mq,Saffari:2007zt,Soroushfar:2015wqa}.

The main purpose of this paper, is to investigate that the
thermodynamic geometric methods can be used to explain thermodynamics
of black holes in $ f(R) $ gravity, and it is organized as follows,
in Sec.~\ref{section2}, we review a static black hole in $ f(R) $
gravity, then we study the thermodynamic behavior and thermodynamic
geometry methods for this black hole, In Sec.~\ref{section3}, also,
we review a static charged black hole in $ f(R) $ gravity and  study
the thermodynamic behavior and thermodynamic geometry methods for
it, In  Sec.~\ref{section4}, we review a rotating charged black hole
in $ f(R) $ gravity, then we investigate the thermodynamic behavior
and thermodynamic geometry methods for it, as well, and final
results are conclude in Sec.~\ref{section5}
\section{STATIC BLACK HOLE IN $f(R)$ GRAVITY}\label{section2}

In this section, we study the field equations for a static black
hole in $f(R)$ gravity. The action depending on the Ricci scalar in
a generic form is:
\begin{equation}\label{action}
S=\frac{1}{2k}\int d^{4}{x}\sqrt{-g}f(R)+S_{m}.
\end{equation}
Varying the action with respect to the metric results in the field
equations as:
\begin{equation}
F(R)R_{\mu\nu}-\frac{1}{2}f(R)g_{\mu\nu}-(\nabla_{\mu}\nabla_{\nu}-g_{\mu\nu}\square)F(R)=kT_{\mu\nu},
\end{equation}
where $ F(R)=\frac{df(R)}{dr} $ and $ \square=\nabla_{\alpha}\nabla^{\alpha}$.

A generic form of the metric of the spherically symmetric spacetime we are considering is
\begin{equation}
ds^{2}=-B(r)dt^{2}+A(r)dr^{2}+r^{2}(d\theta^{2}+\sin^{2}\theta d\varphi^{2}),\quad
\end{equation}
where $A(r)=B(r)^{-1}$.
The model employed for $f(R)$ gravity is given by
\begin{equation}
f(R)=R+\Lambda+\frac{R+\Lambda}{R/R_0 +
2/\alpha}\ln\frac{R+\Lambda}{R_c},
\end{equation}
In which, $R_c$ is a constant of integration and $R_0=6{\alpha}^2/d^2$, where $\alpha$,
and $d$, are free parameters of the action, and also $\Lambda$ is the cosmological constant. The metric solution up to
the first order in the free parameters of the action is obtained as
$B(r)=1-\frac{2m}{r}+\beta r-\frac{1}{3}\Lambda{r}^{2}$, where,
$\beta=\alpha/d\geq0$, is a real constant~\cite{Saffari:2007zt},~\cite{Soroushfar:2015wqa}.\\

\textbf{2.1.} \textbf{Thermodynamic}

In this section, we study the thermodynamic properties of this black
hole. we could find the mass of the black hole $M$, in terms of its
entropy $S$, and the radius of curvature of de Sitter space $ l $,
where $ l $, is related to the cosmological constant $\Lambda$,
through the relation~\cite{Tharanath:2014ika}
\begin{equation}\label{Lambda}
\Lambda=\frac{3}{l^{2}}.
\end{equation}
Using the relation between entropy $S$, and event horizon radius
$r_{+}$, $ (S=\pi r^{2}_{+})$, we can write the mass as below,

\begin{equation}\label{mass}
M(S,l,\beta)=\frac{l^{2}\pi^{\frac{1}{2}}\beta S+l^{2}\pi S^{\frac{1}{2}}-S^{\frac{3}{2}}}{2l ^{2}\pi^{\frac{3}{2}}}.
\end{equation}
The other thermodynamic parameters can be calculated by using the
above expression as Temperature $(T=\frac{\partial M}{\partial S})$
and, heat capacity $(C=T\frac{\partial S}{\partial T})$ as a
function of $S$, $l$ and $\beta$,

\begin{equation}
T=\frac{2\beta l^{2}\pi^{\frac{1}{2}}S^{\frac{1}{2}}-3S+l^{2}\pi}{4l^{2}\pi^{\frac{3}{2}}S^{\frac{1}{2}}},
\end{equation}

\begin{equation}
C=-\frac{4\beta l^{2}\pi^{\frac{1}{2}}S^{\frac{3}{2}}-6S^{2}+2l^{2}\pi S}{l^{2}\pi +3S}.
\end{equation}

\begin{figure}[h]
    \centering
        \includegraphics[width=0.4\textwidth]{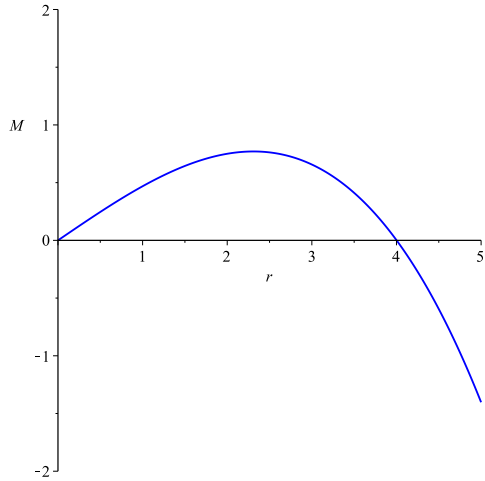}
    \caption{Mass variation of a static black hole in terms of horizon radius $ r_{+} $ for $ l=4.0 $, $ \beta=10^{-4} $.}
 \label{pic:staticM}
\end{figure}

\begin{figure}[h]
    \centering
        \includegraphics[width=0.4\textwidth]{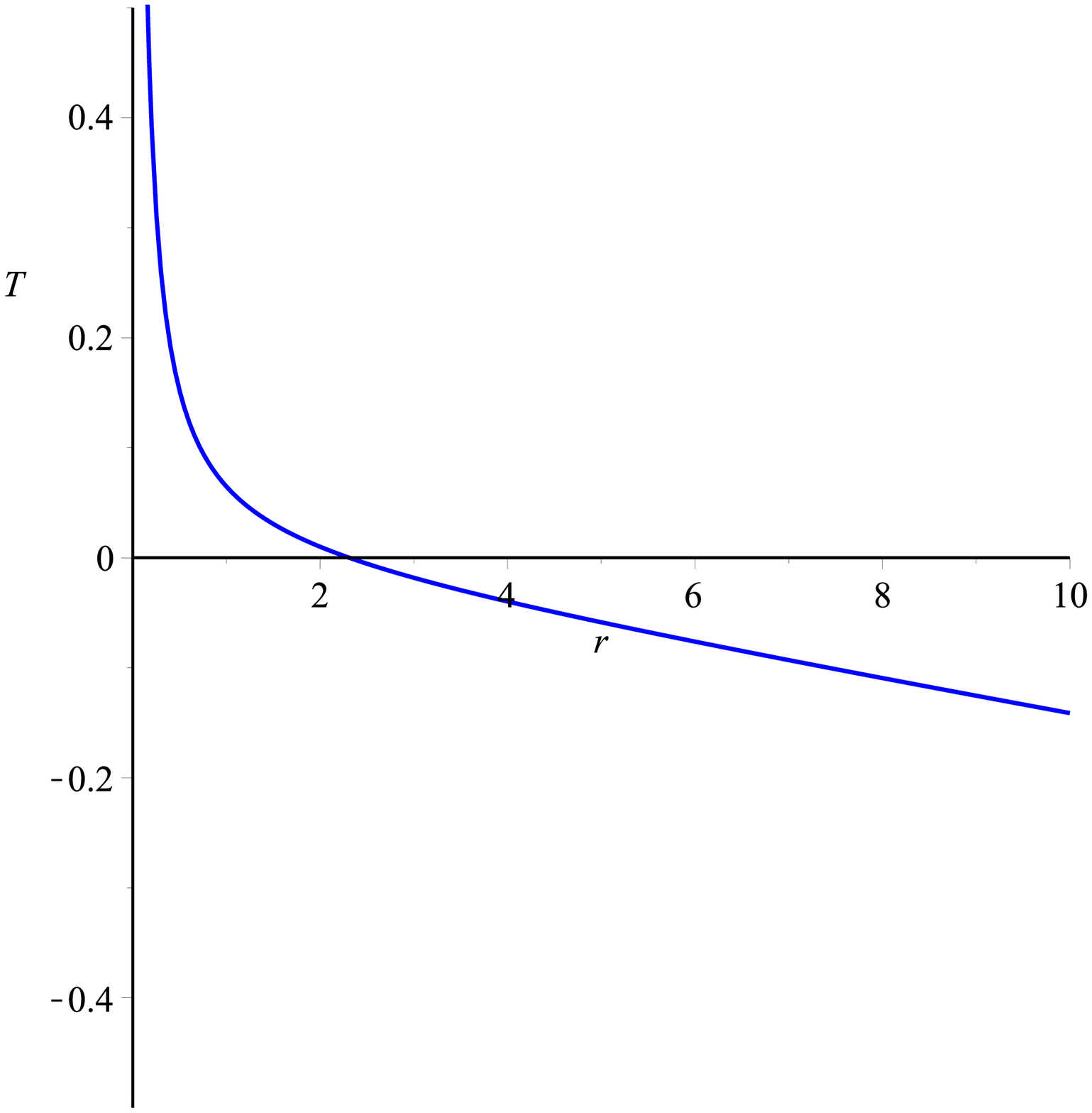}
    \caption{Temprature variation of a static black hole in terms of horizon radius $ r_{+} $ for $ l =4.0 $, $ \beta =10^{-4} $.}
 \label{pic:static T}
\end{figure}

\begin{figure}[h]
    \centering
        \includegraphics[width=0.4\textwidth]{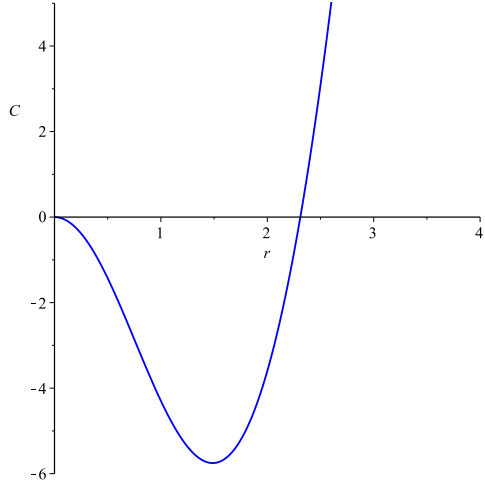}
    \caption{Heat capacity variation of a static black hole in terms of horizon radius $ r_{+} $ for $ l =4.0 $, $ \beta =10^{-4} $.}
 \label{pic:staticC}
\end{figure}

We have obtained three thermodynamic parameters of this black hole
and plotted all of them in terms of horizon radius $r_{+}$, (see
Figs.~\ref{pic:staticM}--\ref{pic:staticC}).
\\

In Fig.\ref{pic:staticM}, it can be seen that, mass of the black
hole become zero at two points, $r_{+}=r_{01}$ and $r_{+}=r_{02}$ (we show
the zero points of mass with $ r_{01} $ and $ r_{02} $), in which, $
r_{01}=0 $ and $ r_{02}=4.0 $, and it reaches into a maximum value
at $r_{+}=r_{m}$ (we show the place of maximum value of mass with $
r_{m} $), which is equal to 2.31. Also, it can be observed from
Fig.~\ref{pic:static T}, that, Temperature is positive only in a
particular range of $r_{+}$, then it reaches in to zero at,
$r_{+}=r_{m}$, and after that, it falls in to negative region, in
which, it has nonphysical meaning. Finally, by plotting heat capacity
of the black hole in terms of horizon radius, $r_{+}$, in
Fig.~\ref{pic:staticC}, we have shown that, this black hole has
phase transition type one, in other words, in the range of, $
0<r_{+}<r_{m} $, the heat capacity is in the negative region
(unstable phase), then at, $ r_{+}=r_{m} $, it takes phase transition
type one ($ C(r_{+}=r_{m})=0) $, after that for, $ r_{+}>r_{m} $, it
will be positive (stable).
\\

\textbf{2.2.} \textbf{Thermodynamic Geometry}

Now, we construct the geometric structure for this black hole by
applying the geometric technique of Weinhold, Ruppiner and GTD
metrics of the system. In this case, the extensive variables are,
$N^{r}=(l, \beta)$. According to Eq.~(\ref{Weinhold}), we can write
the Weinhold metric for this system as below
 \begin{equation}
 g^{W}_{i j}=\partial _{i}\partial _{j}M(S,l, \beta),
 \end{equation}

\begin{eqnarray}
ds^{2}_{W}=M_{SS}dS^{2}+M_{l l}dl^{2}+M_{\beta\beta}d\beta^{2}\nonumber\\
  2M_{Sl}dSdl +2M_{S\beta}dSd\beta +2M_{l\beta}dl d\beta ,
\end{eqnarray}

 therefore
\begin{equation}
g^{W}=\begin{bmatrix}
M_{SS} & M_{Sl} & M_{S\beta}\\
M_{l S} & M_{l \alpha} &0\\
M_{\beta S} & 0 & 0
\end{bmatrix}.
\end{equation}

The components of above matrix can be found using the expression of
$M$, given in Eq.~(\ref{mass}). We could calculate the curvature
scalar of the Weinhold metric as,
\begin{equation}
R^{W}=0,
\end{equation}
so the Weinhold structure is flat for this black hole and, we can not
explain phase transition of this thermodynamic system. Now, we use
Ruppiner method, which is conformaly transformed to Weinhold
metric. Ruppiner metric is given by

\begin{equation}
ds^{2}_{R}=\frac{1}{T}ds^{2}_{W}.
\end{equation}
The correspond matrix with the metric components of Ruppiner method, is as follow,

\begin{equation}
g^{R}=(\frac{1}{T})\begin{bmatrix}
M_{SS} & M_{Sl} & M_{S\beta}\\
M_{l S} & M_{ll} &0\\
M_{\beta S} & 0 & 0
\end{bmatrix},
\end{equation}
which is equal to
\begin{equation}
g^{R}=(\frac{4l^{2}\pi^{\frac{3}{2}}S^{\frac{1}{2}}}{2l^{2}\pi^{\frac{1}{2}}S^{\frac{1}{2}}\beta -3S+l^{2}\pi})\begin{bmatrix}
M_{SS} & M_{Sl} & M_{S\beta}\\
M_{l S} & M_{ll} &0\\
M_{\beta S} & 0 & 0
\end{bmatrix}.
\end{equation}

\begin{figure}[h]
    \centering
        \includegraphics[width=0.4\textwidth]{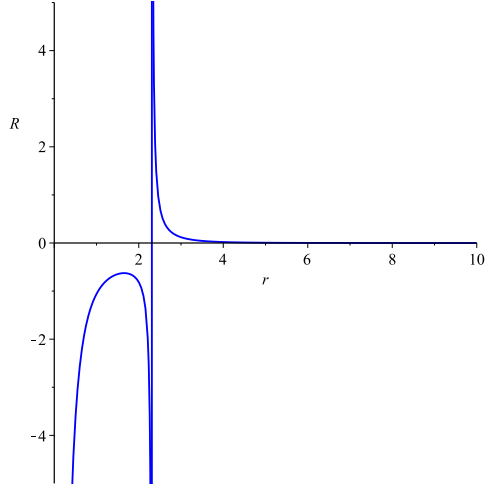}
    \caption{Variation of Ruppiner metric in terms of horizon radius $ r_{+} $ for $l =4.0$, $\beta =10^{-4}$.}
 \label{pic:staticR}
\end{figure}

The curvature of the Ruppiner metric is obtained as below,

\begin{equation}
R^{R}=\frac{4S^{\frac{5}{2}}l^{2}\pi^{\frac{1}{2}}\beta -13S^{\frac{5}{2}}\pi^{\frac{1}{2}}l^{2}\beta -3S^{\frac{5}{2}}\pi^{\frac{1}{2}}l^{2}\beta -11S^{2}l^{2}\pi +3S^{3}}{4S^{3}(2S^{\frac{1}{2}}\pi^{\frac{1}{2}}l^{2}\beta -3S+l^{2}\pi)},
\end{equation}
which is singular at $S=0$ and, $S=\frac{1}{3}l^{2}\pi (2\beta l
(\frac{1}{3}l\beta +\frac{1}{3}\sqrt{l^{2}\beta^{2}+3})+1)$, for
each solution of $ S $, there exists a pair of $r_{+}$,
$(r_{+}=\pm\sqrt{\frac{S}{\pi}})$, which can explain zero points in
this thermodynamic system. We avoid the negative values of this
solution because it gives imaginary and negative roots. The values of
these zero points are, $r_{+}=0$, and, $r_{+}=r_{m}$. It is
completely coincide with zero point of the temperature and the heat capacity
(the phase transition point) of this black hole. The curvature scalar of Ruppeiner
metric for this black hole with respect to horizon radius, $ r_{+} $,
is demonstrated in Fig.~\ref{pic:staticR}.
\\

\begin{figure}[h]
    \centering
        \includegraphics[width=0.4\textwidth]{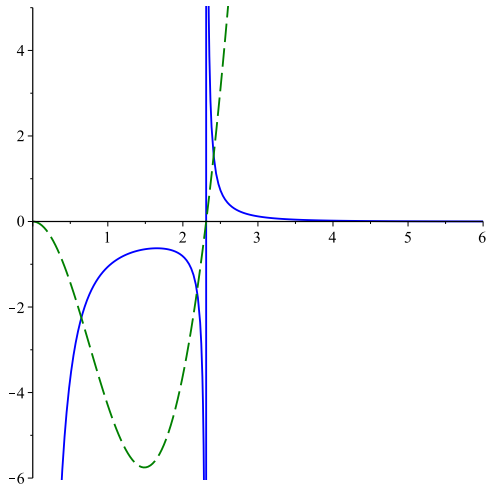}
    \caption{Curvature scalar variation of Ruppeiner metric (blue continuous line) and the heat capacity of a static black hole (gree dash line) in terms of horizon radius $ r_{+} $, for $l =4.0$, $\beta =10^{-4}$.}
 \label{pic:staticRC}
\end{figure}

Plot of scalar curvature of Ruppeiner metric and heat capacity, in
terms of, $ r_{+} $, have shown in Fig.~\ref{pic:staticRC}. Also, it
can be seen from Figs.~\ref{pic:staticC}--\ref{pic:staticRC}, that,
singular points of scalar curvature are coincide with zero point of
heat capacity.

Finally, we construct the most important metric in GTD method, in
which the choice of thermodynamic potential is not important, the metric
for this thermodynamic system according to Eq.~(\ref{GTD}), is as follows

\begin{equation}
g^{GTD}=(SM_{S}+l M_{l}+\beta M_{\beta})\begin{bmatrix}
-M_{SS} & 0 & 0\\
0 & M_{ll} & 0\\
0 & 0 & 0
\end{bmatrix}.
\end{equation}

We can not obtain the corresponding curvature scalar with this metric,
because, the metric determinant is zero, so, inverse of the metric is infinite, therefore; 
in this case we can not find any
physical information about the system from the GTD method.

\begin{figure}[h]
    \centering
     \subfigure[]{
        \includegraphics[width=0.4\textwidth]{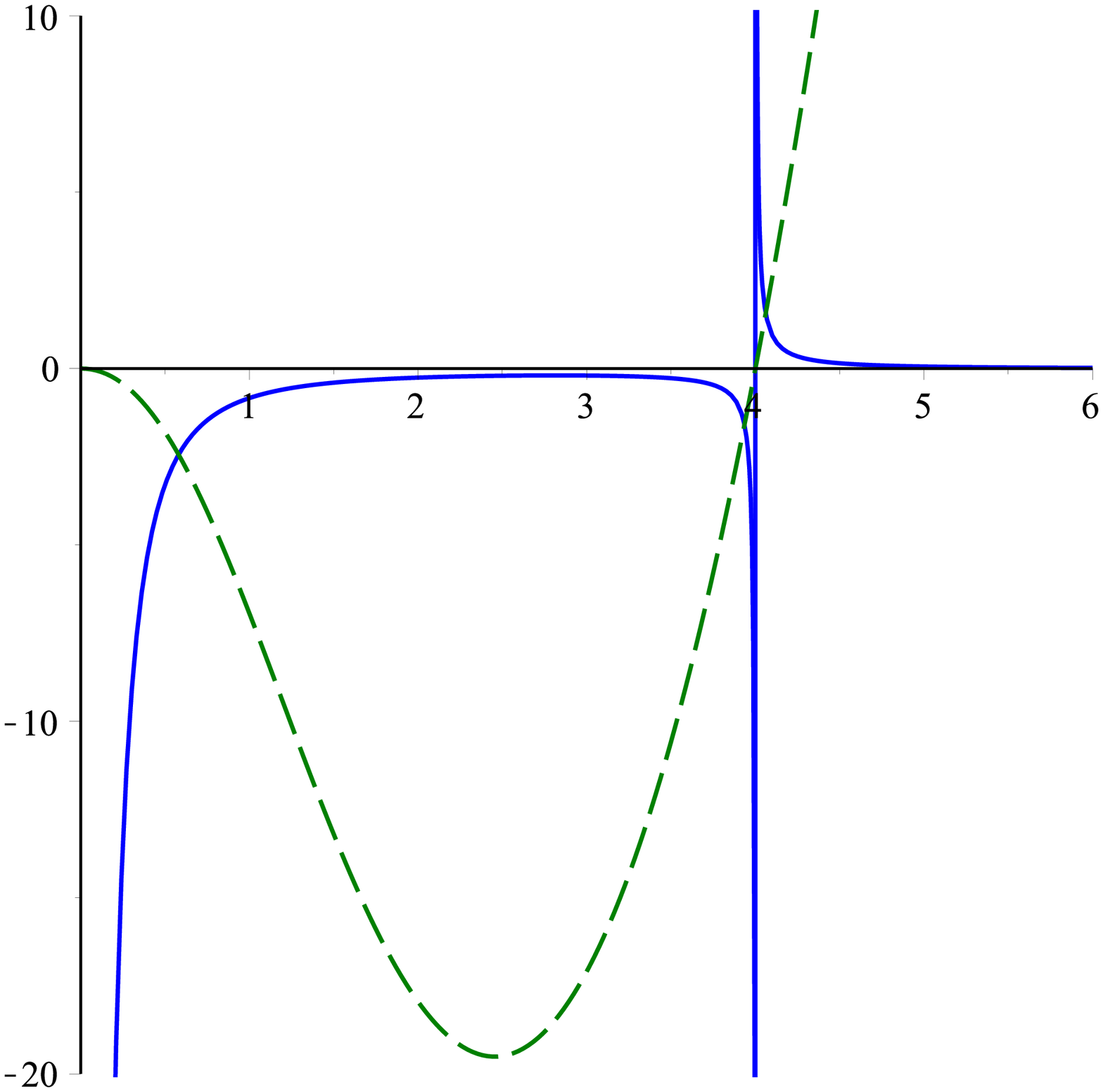}
    }
    \subfigure[]{
        \includegraphics[width=0.4\textwidth]{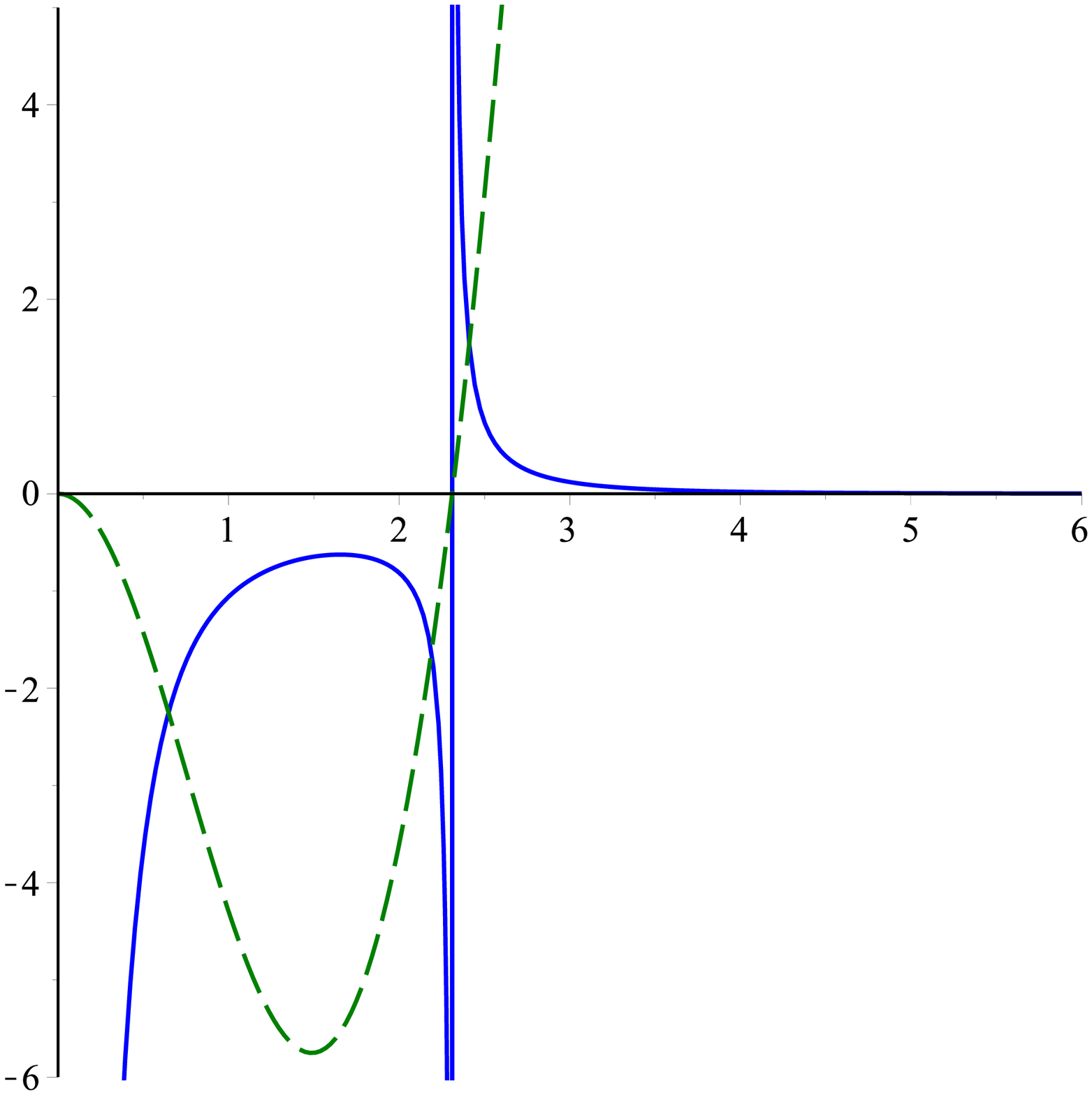}
    }
    \subfigure[]{
        \includegraphics[width=0.4\textwidth]{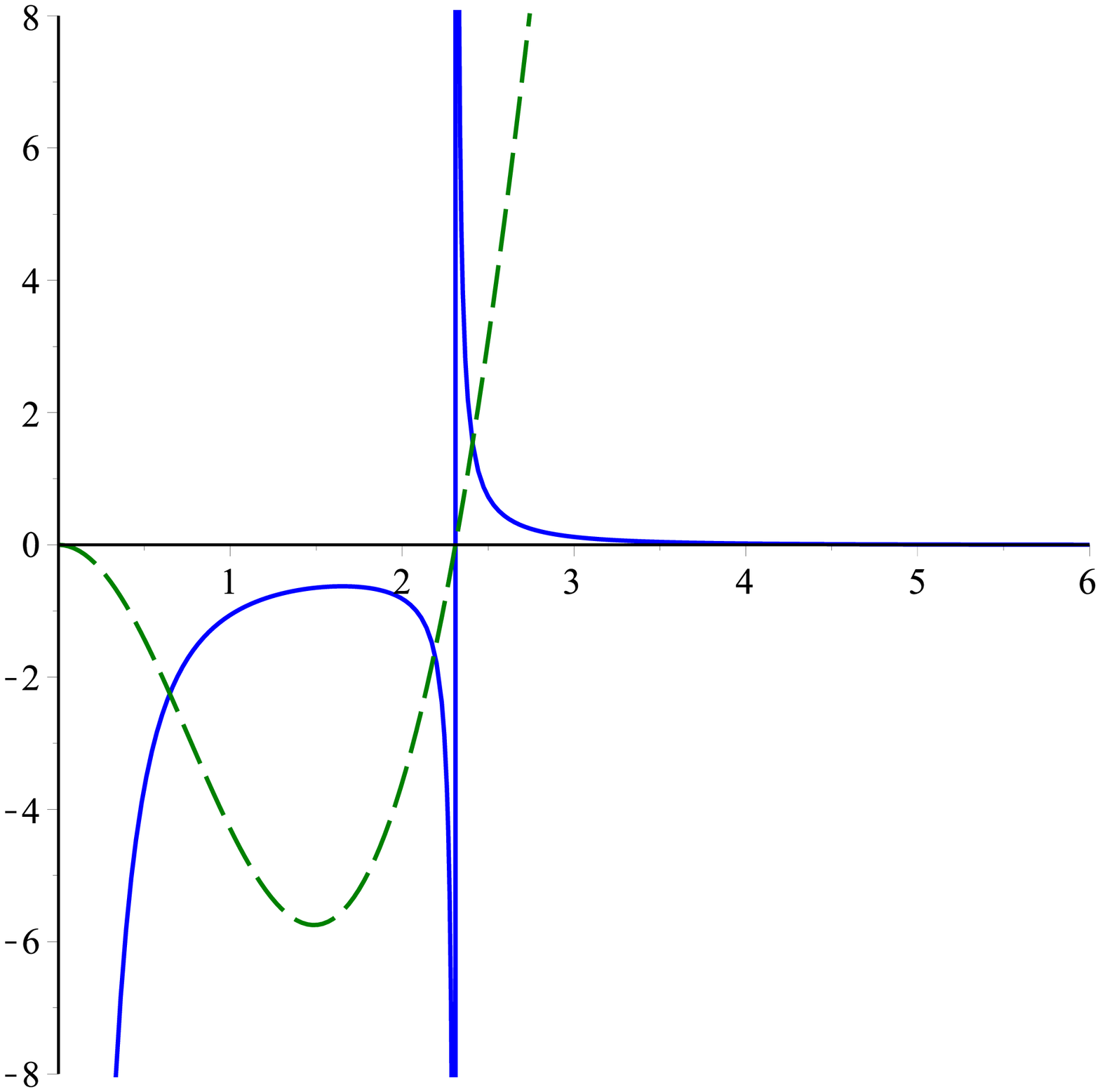}
    }
    \caption{ Curvature scalar variation of Ruppeiner metric (blue continuous line) and the heat capacity of a static black hole (gree dash line) in terms of $ r_{+} $, for $ l=4.0 $ and $ \beta=0.25 $, $ \beta=10^{-4}$, $\beta=10^{-15} $, for (a), (b) and (c), respectively.}
 \label{pic:CSb}
\end{figure}

\begin{figure}[h]
    \centering
     \subfigure[]{
        \includegraphics[width=0.4\textwidth]{static/CSb1.eps}
    }
    \subfigure[]{
        \includegraphics[width=0.4\textwidth]{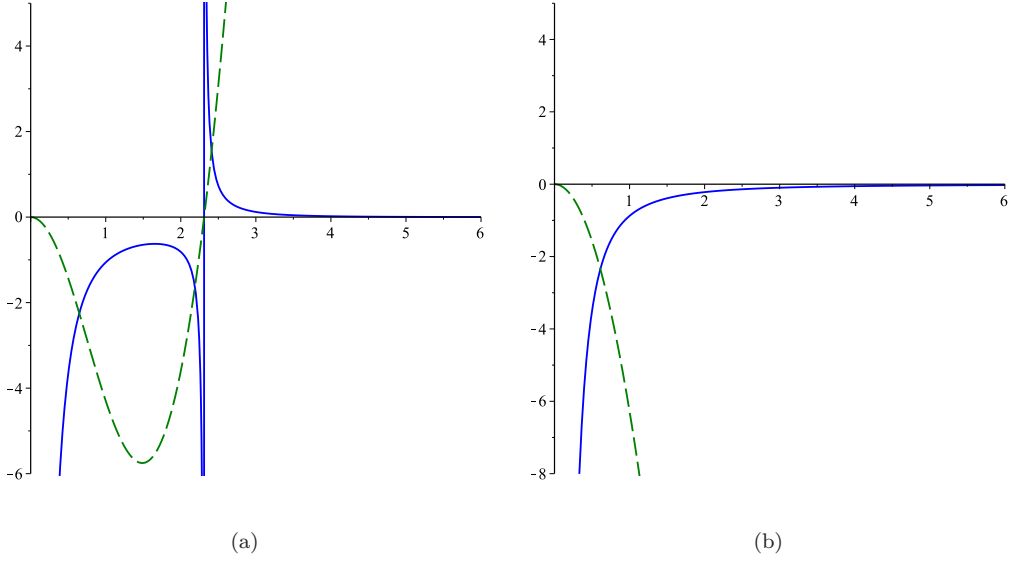}
    }
    \caption{Curvature scalar variation of Ruppeiner metric (blue continuous line) and the heat capacity of a static black hole (gree dash line) in terms of $ r_{+} $ for $ \beta=10^{-4} $, and $ l=4.0 $, $ l=\sqrt{3\cdot 10^{15}} $, for (a) and (b), respectively.}
 \label{pic:CSb1}
 \end{figure}

Now, at the end of this section, we investigate the effect of
changes in the value of $ \beta $, and $ l $, parameters on phase
transition points. It is clear from Fig.~\ref{pic:CSb}, by
decreasing value of $ \beta $, we do not have any changing in number
of phase transition, but the place of it will decrease. In
Fig.~\ref{pic:CSb1}(a), we find that, for small value of $ l $, the
system has phase transition type one, but for the large
value of $ l $, the system is in the unstable phase and it has no
phase transition (see Fig.~\ref{pic:CSb1}(b)).
\\

\clearpage

\clearpage

In the next section, we investigate the static charged black hole in $f(R)$ gravity.

\section{STATIC CHARGED BLACK HOLE IN $f(R)$ GRAVITY}\label{section3}

In this section, we describe metric and the field equations of
a static charged black hole in $f(R)$ gravity. Here, the action for $
f(R) $ gravity, with Maxwell term in four dimensions, is

\begin{equation}\label{action}
S=\frac{1}{16\pi G}\int d^{4}{x}\sqrt{-g}(R+f(R)-F_{\mu\nu}F^{\mu\nu}).
\end{equation}
Varying the action with respect to the metric results in the field equations as:
\begin{align}\label{rmiyonu}
R_{\mu\nu}\big(1+ f'(R)\big)-\frac{1}{2}\big(R+f(R)\big)g_{\mu\nu}
+\big(g_{\mu\nu}\nabla^{2}-\nabla_{\mu}\nabla_{\nu}\big)f'(R)=2T_{\mu\nu},
\end{align}
where $ T_{\mu\nu} $, is the stress-energy tensor of the
electromagnetic field, which is given by

\begin{equation}
T_{\mu\nu}=F_{\mu\rho}F_{\nu}^{\rho}-\dfrac{g_{\mu\nu}}{4}F_{\rho\sigma}F^{\rho\sigma},
\end{equation}
with
\begin{equation}
T^{\mu} _{\,\,\mu}=0 ,
\end{equation}
$R_{\mu\nu}$ is the Ricci tensor, and $\nabla$ is the usual covariant derivative.
The trace of Eq.~(\ref{rmiyonu}), for $R=R_{0}$, yields,
\begin{equation}
R_{0}\big(1+f'(R_{0})\big)-2\big(R_{0}+f(R_{0})\big)=0,
\end{equation}
which determines the negative constant curvature scalar as
\begin{equation}\label{R0}
R_{0}=\dfrac{2f(R_{0})}{f'(R_{0})-1}.
\end{equation}
Using Eqs.~(\ref{rmiyonu})--(\ref{R0}), the Ricci tensor is
\begin{align}
R_{\mu\nu}=\dfrac{1}{2}\big(\dfrac{f(R_{0})}{f'(R_{0})-1}\big)g_{\mu\nu}
+\dfrac{2}{\big(1+f'(R_{0})\big)}T_{\mu\nu}.
\end{align}
Finally, the metric of the spherically symmetric spacetime is given by
\begin{equation}
ds^{2}=N(r)dt^{2}-N(r)^{-1}dr^{2}-r^{2}(d\theta^{2}+\sin^{2}\theta d\varphi^{2}),\quad
\end{equation}
with
\begin{equation}\label{N}
N(r)=1-\frac{2GM}{r}+\frac{Q^{2}}{(1+f^{'}(R_{o}))r^{2}}-\frac{1}{12}R_{0}{r}^{2}.
\end{equation}
For a general discussion of this metric, see Ref.~\cite{Moon:2011hq}.
In the following, we consider, $ G=1 $, and
$ q^{2}=\frac{Q^{2}}{(1+f^{'}(R_{o}))} $, therefore we have
\begin{equation}\label{metric}
ds^{2}=(1-\frac{2M}{r}+\frac{q^{2}}{r^{2}}-\frac{1}{12}R_{0}{r}^{2})dt^{2}-(1-\frac{2M}{r}+\frac{q^{2}}{r^{2}}-\frac{1}{12}R_{0}{r}^{2})^{-1}dr^{2}-r^{2}(d\theta^{2}+\sin^{2}\theta d\varphi^{2}),\quad
\end{equation}
where $ R_{0}=4\Lambda $, in which $ \Lambda $ is the cosmological constant, and $ q $, is the electrical charge.\\

\textbf{3.1.} \textbf{Thermodynamic}

Now, in this section we investigate the thermodynamic properties of
this black hole. By solving Eq.(\ref{N}) in terms of $r_{+}$
$(N(r_{+})=0)$ and, using the relation between entropy $S$, and
horizon radius $r_{+}$, the mass of this black hole will be obtained
in terms of the entropy, charge and the radius of de Sitter space,
as below

\begin{equation} \label{mass1}
M(S,l,q)=\frac{l^{2}\pi^{2}q^{2}+l^{2}\pi S-S^{2}}{2l^{2}\pi^{\frac{3}{2}}S^{\frac{1}{2}}}.
\end{equation}

\begin{figure}[h]
    \centering
        \includegraphics[width=0.4\textwidth]{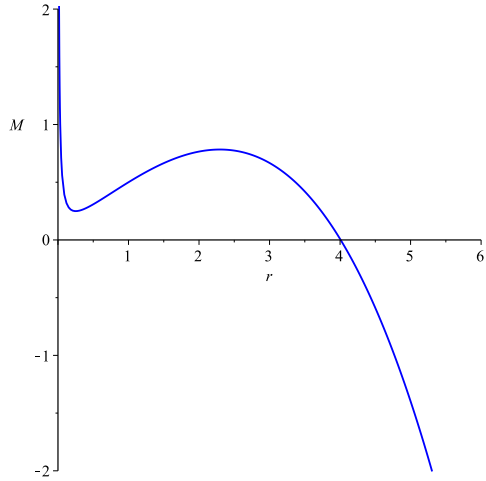}
    \caption{Mass variation of a charged static black hole in terms of horizon radius $ r_{+} $ for $ q=0.25 $, $ l=4.0 $.}
 \label{pic:chargedM}
\end{figure}

\begin{figure}[h]
    \centering
        \includegraphics[width=0.4\textwidth]{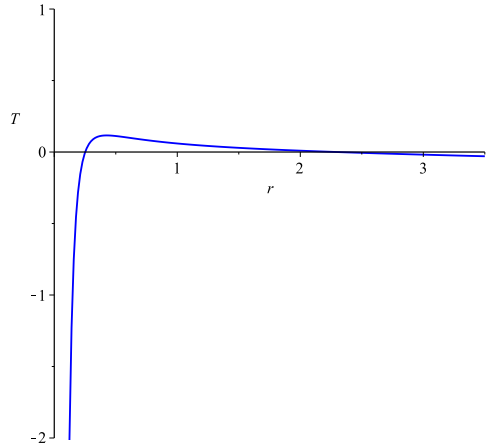}
    \caption{Temperature variation of a charged static black hole in terms of horizon radius $ r_{+} $ for $ q=0.25 $, $ l=4.0 $.}
 \label{pic:chargedT}
\end{figure}

\begin{figure}[h]
    \centering
     \subfigure[]{
        \includegraphics[width=0.4\textwidth]{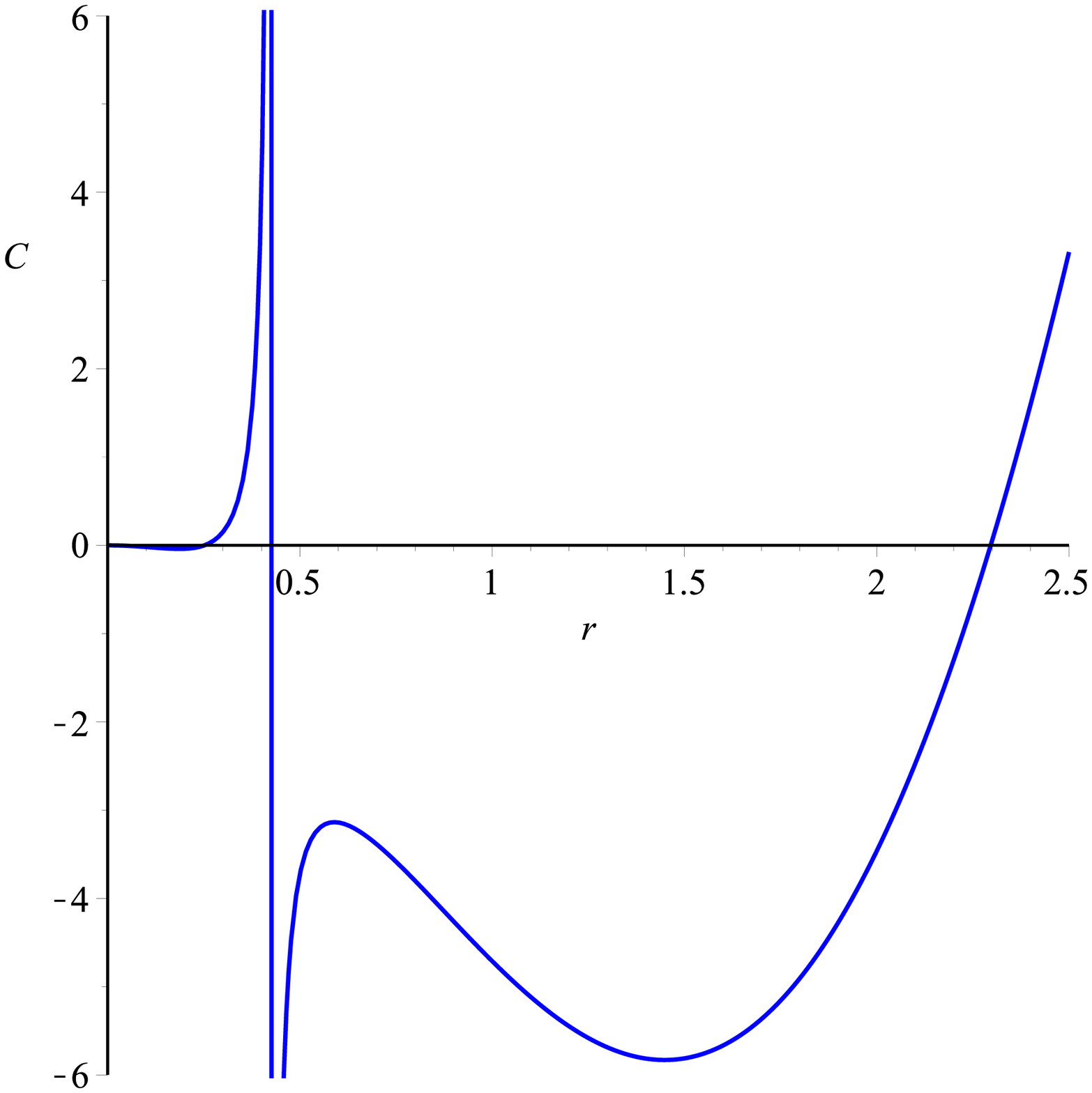}
    }
    \subfigure[Closeup of figure (a)]{
        \includegraphics[width=0.4\textwidth]{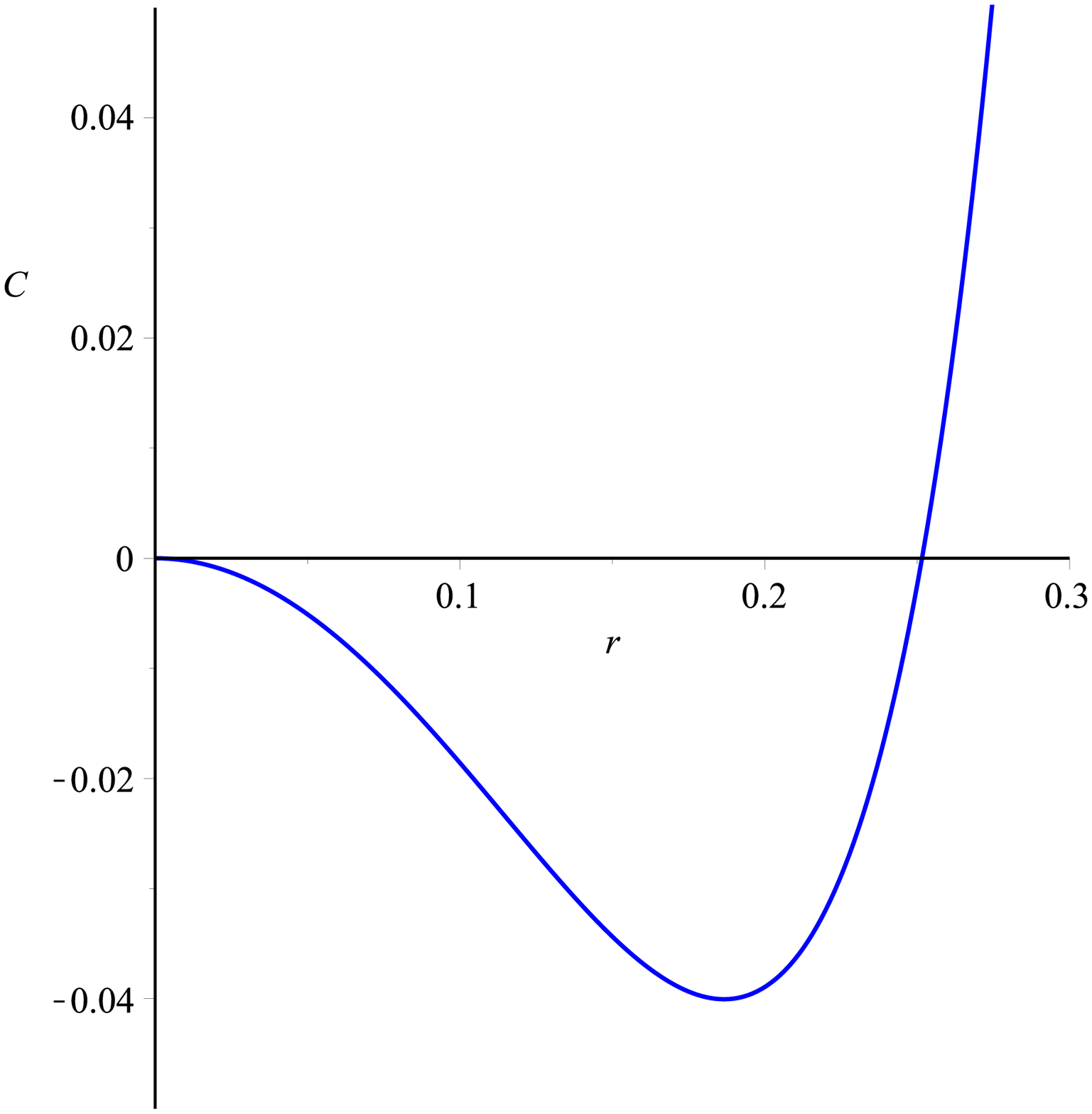}
    }
    \caption{Heat capacity variation of a charged static black hole in terms of horizon radius, $ r_{+} $ for $ l=4.0 $, $ q=0.25 $.}
 \label{pic:chargedC}
\end{figure}

In following, we can straightforwardly write the temperature,
the electrical potential and the heat capacity of the black hole, from
the first low of thermodynamic as follows

\begin{equation}
dM=TdS+\Phi dq,
\end{equation}

\begin{equation}\label{temp}
T=\frac{l^{2}\pi S-l^{2}\pi^{2}q^{2}-3S^{2}}{4l^{2}\pi^{\frac{3}{2}}S^{\frac{3}{2}}},
\end{equation}

\begin{equation}
\Phi =q\sqrt{\frac{\pi}{S}} ,
\end{equation}

\begin{equation}\label{heat}
C=\frac{2l^{2}\pi^{2}q^{2}S-2l^{2}\pi S^{2}+6S^{3}}{-3l^{2}\pi^{2}q^{2}+l^{2}\pi S+3S^{2}}.
\end{equation}
The plots of Eqs.~(\ref{mass1})--(\ref{heat}) are demonstrated in Figs.~\ref{pic:chargedM}--\ref{pic:chargedC}.
\\

\clearpage

Fig.~\ref{pic:chargedM}, shows that, the mass of this black hole has
a minimum value at $ r_{+}=r_{m1} $, (we show the minimum point of
the mass with $ r_{m1} $), which has the value equal to 0.252, then
it reaches to its maximum value at $ r_{+}=r_{m2} $ (we show the
maximum point of mass with $ r_{m2} $), which its value is equal to
2.296, and it vanishes at $ r=r_{0} $, $ r_{0} $ is the point that,
mass becomes zero, and it is equal to 4.0 . This is also observed
from Fig.~\ref{pic:chargedT}, that, temperature is positive only
in a particular range of event horizon $ (r_{m1}<r_{+}<r_{m2}) $, in
addition, at $ r_{+}<r_{m1} $, and $ r_{m2}<r_{+} $, it will be
negative and, it has no physical solution. As well as, it can be
observed from Fig.~\ref{pic:chargedC}, that, the heat capacity of
this black hole will be zero at $ r_{m1} $, and $ r_{m2} $, ($
C(r_{+}=r_{m1})=0$ and $C(r_{+}=r_{m2})=0 $), in the other words, it
has two phase transition type one at these points, moreover, at
$ r_{+}=r_{\infty} $ (we show the divergence point of heat capacity
with $ r_{\infty} $), heat capacity diverges, that, the value of
this point is equal to 0.426. In other words, at $ r_{+}<r_{m1} $,
heat capacity is negative and it is in unstable phase, then, at $
r_{m1}<r_{+}<r_{\infty} $, heat capacity is positive or it is in
stable phase, afterward, at $ r_{\infty}<r_{+}<r_{m2} $, it falls in
to negative region (unstable phase) and, at $ r_{+}>r_{m2} $, it
becomes stable.
\\

\textbf{3.2.} \textbf{Thermodynamic Geometry}

In this section, we construct the thermodynamic geometry structure
for this black hole. First, we use Weinhold method. Extensive
variables for this system are $ N^{r}=(l, q) $, so the resulting
matrix of the Weinhold metric becomes

\begin{equation}
g^{W}=\begin{bmatrix}
M_{SS} & M_{Sq} & M_{Sl}\\
M_{qS} & M_{qq} &0\\
M_{l S} & 0 & M_{ll}
\end{bmatrix}.
\end{equation}

The elements of the metric can be obtained from the
Eq.~(\ref{mass1}), and  Weinhold scalar curvature can be found as

\begin{equation}
R^{W}=\frac{9l^{2}\pi^{\frac{3}{2}}S^{\frac{5}{2}}-l^{4}\pi^{\frac{7}{2}}q^{2}S^{\frac{1}{2}}-l^{4}\pi^{\frac{5}{2}}S^{\frac{3}{2}}}{(l^{2}\pi^{2}q^{2}-l^{2}\pi S+3S^{2})^{2}}.
\end{equation}

\begin{figure}[h]
    \centering
        \includegraphics[width=0.4\textwidth]{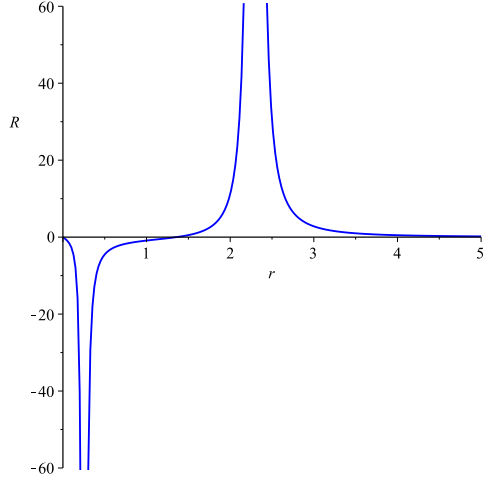}
    \caption{Curvature scalar variation of the Weinhold metric in terms of horizon radius $ r_{+} $ for $ l=4.0 $, $ q=0.25 $.}
 \label{pic:chargedRW}
\end{figure}

Denominator of the above expression becomes zero at $ S=\frac{\pi l}{6}(l \pm\sqrt{l^{2}-12q^{2}}) $, or at $r_{+}=r_{m1} $,
and $ r_{+}=r_{m2} $, (see, Fig.~\ref{pic:chargedRW}).\\

Next, we use the Ruppeiner method for this black hole. Using Eq.~(\ref{mass1}), matrix components of Ruppeiner
metric will be obtained as

\begin{equation}\label{GR}
g^{R}=\frac{1}{T}\begin{bmatrix}
M_{SS} & M_{Sq} & M_{Sl}\\
M_{qS} & M_{qq} &0\\
M_{l S} & 0 & M_{ll}
\end{bmatrix}.
\end{equation}

So, using Eqs.~(\ref{temp}) and (\ref{GR}), we obtain

\begin{equation}
g^{R}=\frac{4l^{2}\pi^{\frac{3}{2}}S^{\frac{3}{2}}}{l^{2}\pi S-l^{2}\pi^{2}q^{2}-3S^{2}}\begin{bmatrix}
M_{SS} & M_{Sq} & M_{Sl}\\
M_{qS} & M_{qq} &0\\
M_{l S} & 0 & M_{ll}
\end{bmatrix}.
\end{equation}

After some calculation, the corresponding curvature scalar will be obtained as

\begin{equation}
R^{R}=-\frac{l^{2}\pi (2\pi q^{2}-S)}{S(l^{2}\pi^{2}q^{2}-l^{2}\pi S+3S^{2})},
\end{equation}
\\

\begin{figure}[h]
    \centering
        \includegraphics[width=0.4\textwidth]{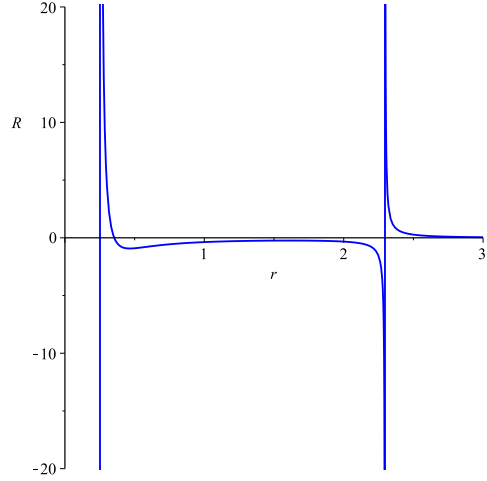}
    \caption{Curvature scalar variation of the Ruppeiner metric in terms of horizon radius $ r_{+} $ for $ l=4.0 $, $ q=0.25 $.}
 \label{pic:chargedRR}
\end{figure}

This curvature scalar is singular at $ S=0 $ and $ S=\frac{\pi l}{6}(l
\pm\sqrt{l^{2}-12q^{2}}) $, or as it can be seen from
Fig.~\ref{pic:chargedRR}, it is singular at $ r_{+}=0 $, $ r_{+}=r_{m1} $ and $
r_{+}=r_{m2} $.

Finally, we use the most important GTD metric. The resulting matrix from the metric is as follows

\begin{equation}
g^{GTD}=(SM_{S}+qM_{q}+lM_{l})\begin{bmatrix}
-M_{SS} & 0 & 0\\
0 & M_{qq} &0\\
0 & 0 & M_{ll}
\end{bmatrix}.
\end{equation}
The corresponding curvature scalar will be obtained as

\begin{equation}
R^{GTD}=\frac{16}{3}\frac{\mathcal{N}}{(3l^{2}\pi^{2}q^{2}-l^{2}\pi S-3S^{2})^{2}(3l^{2}\pi^{2}q^{2}+l^{2}\pi S+S^{2})^{3}}.
\end{equation}

\begin{figure}[h]
    \centering
        \includegraphics[width=0.4\textwidth]{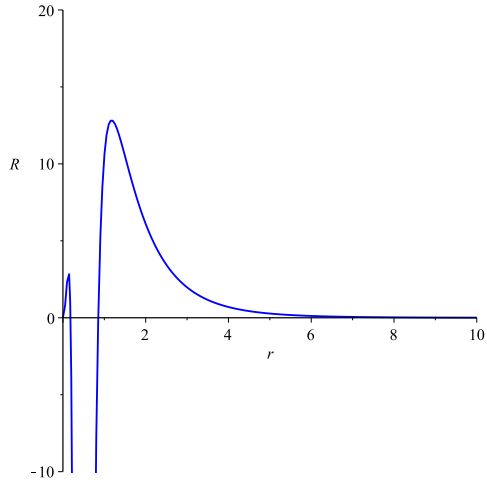}
    \caption{Curvature scalar variation of GTD metric in terms of horizon radius $ r_{+} $ for $ l=4.0 $, $ q=0.25 $.}
 \label{pic:chargedRGTD}
\end{figure}

\begin{figure}[h]
    \centering
        \includegraphics[width=0.4\textwidth]{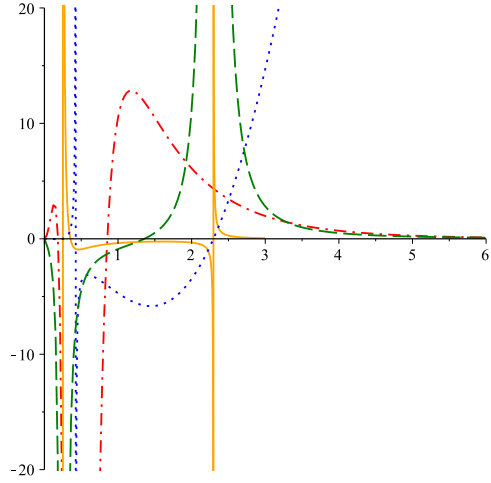}
    \caption{curvature scalar variation of GTD (red dash-dot line), Ruppeiner(orange continuous line), Weinhold (green dash line) metrics, and the heat capacity (blue dot line), in terms of horizon radius $ r_{+} $, for $ l=4.0 $, $ q=0.25 $.}
 \label{pic:chargedRGWC}
\end{figure}

\begin{figure}[h]
    \centering
     \subfigure[]{
        \includegraphics[width=0.4\textwidth]{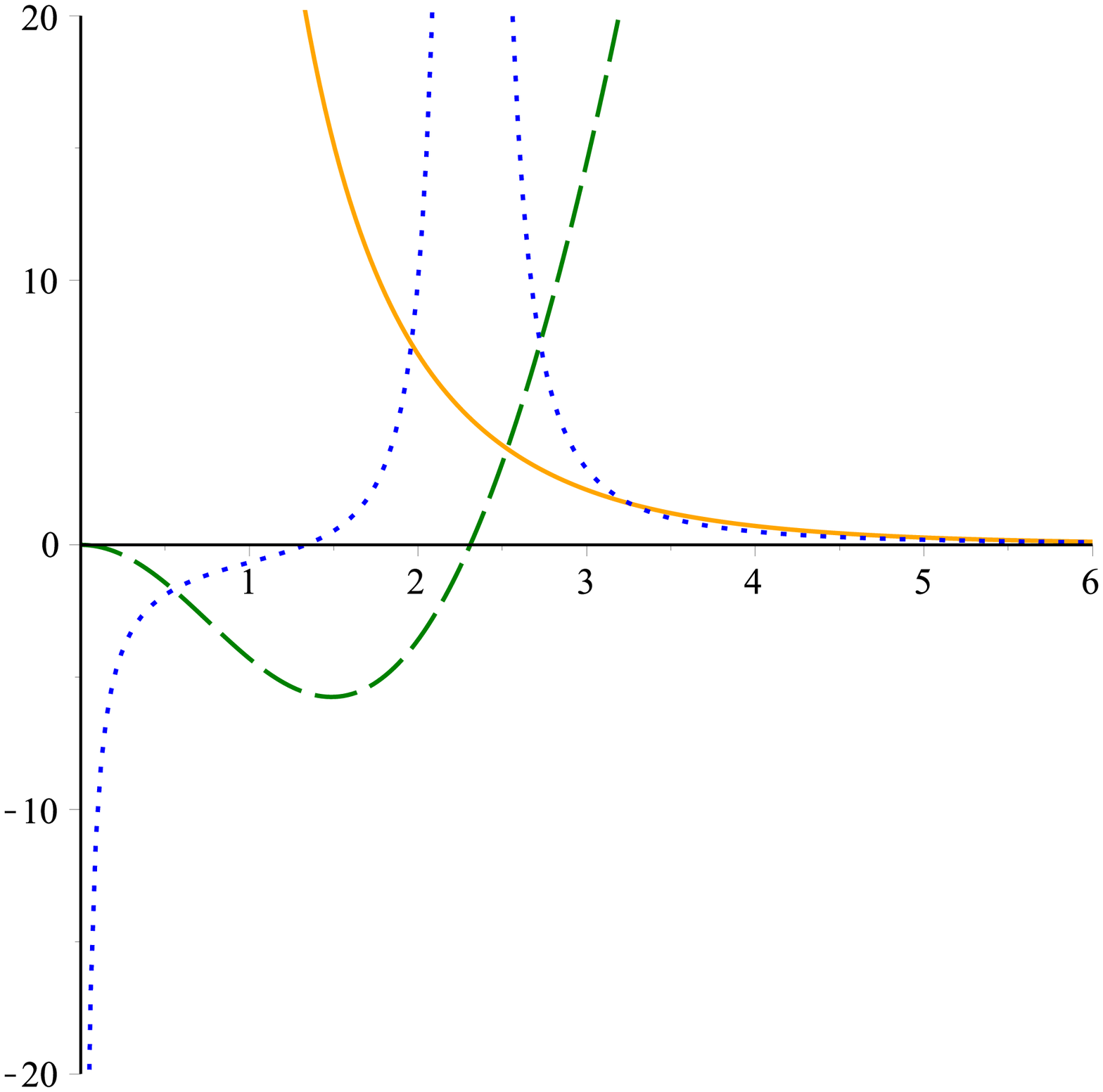}
    }
    \subfigure[]{
        \includegraphics[width=0.4\textwidth]{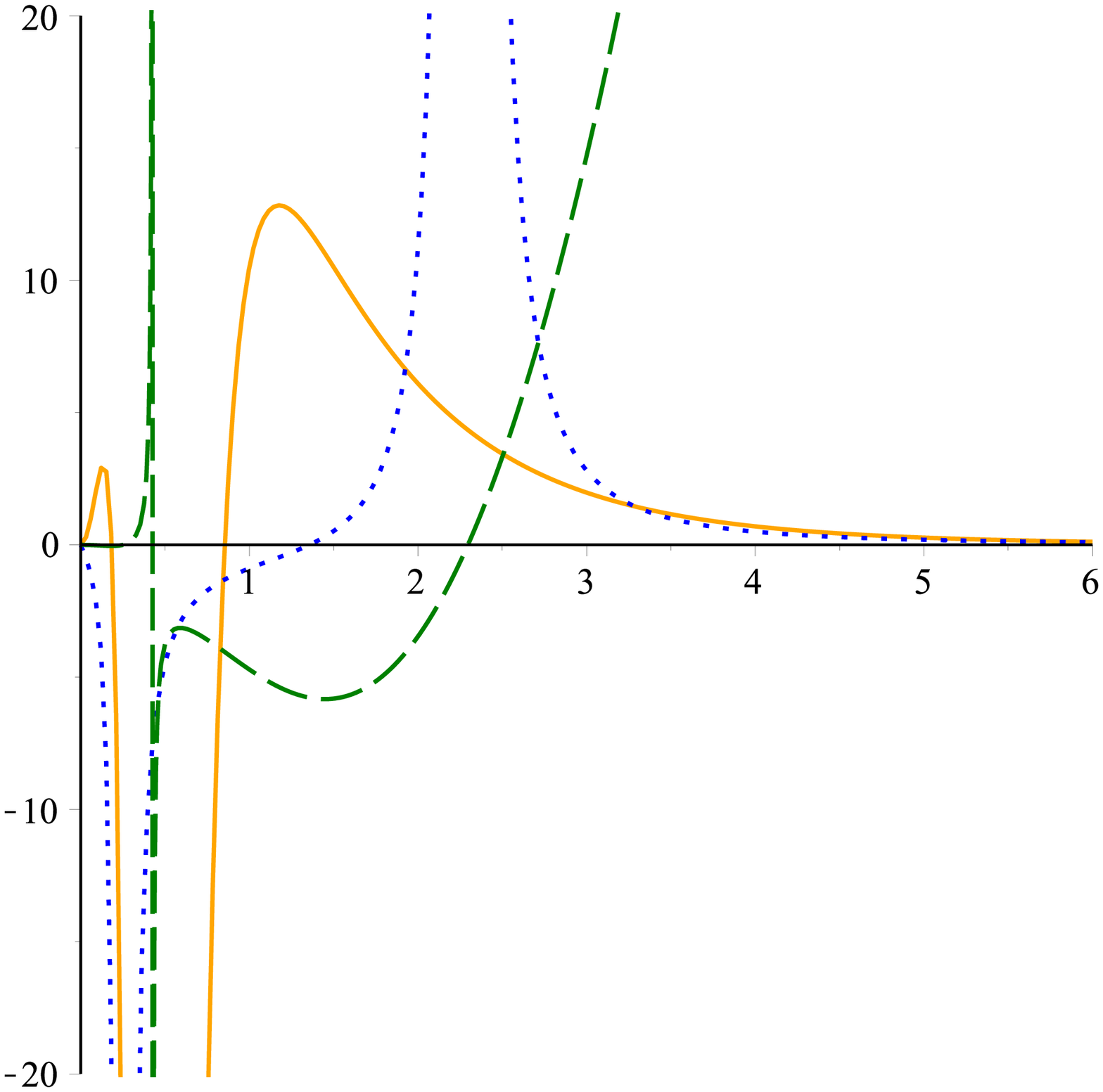}
    }
    \subfigure[]{
        \includegraphics[width=0.4\textwidth]{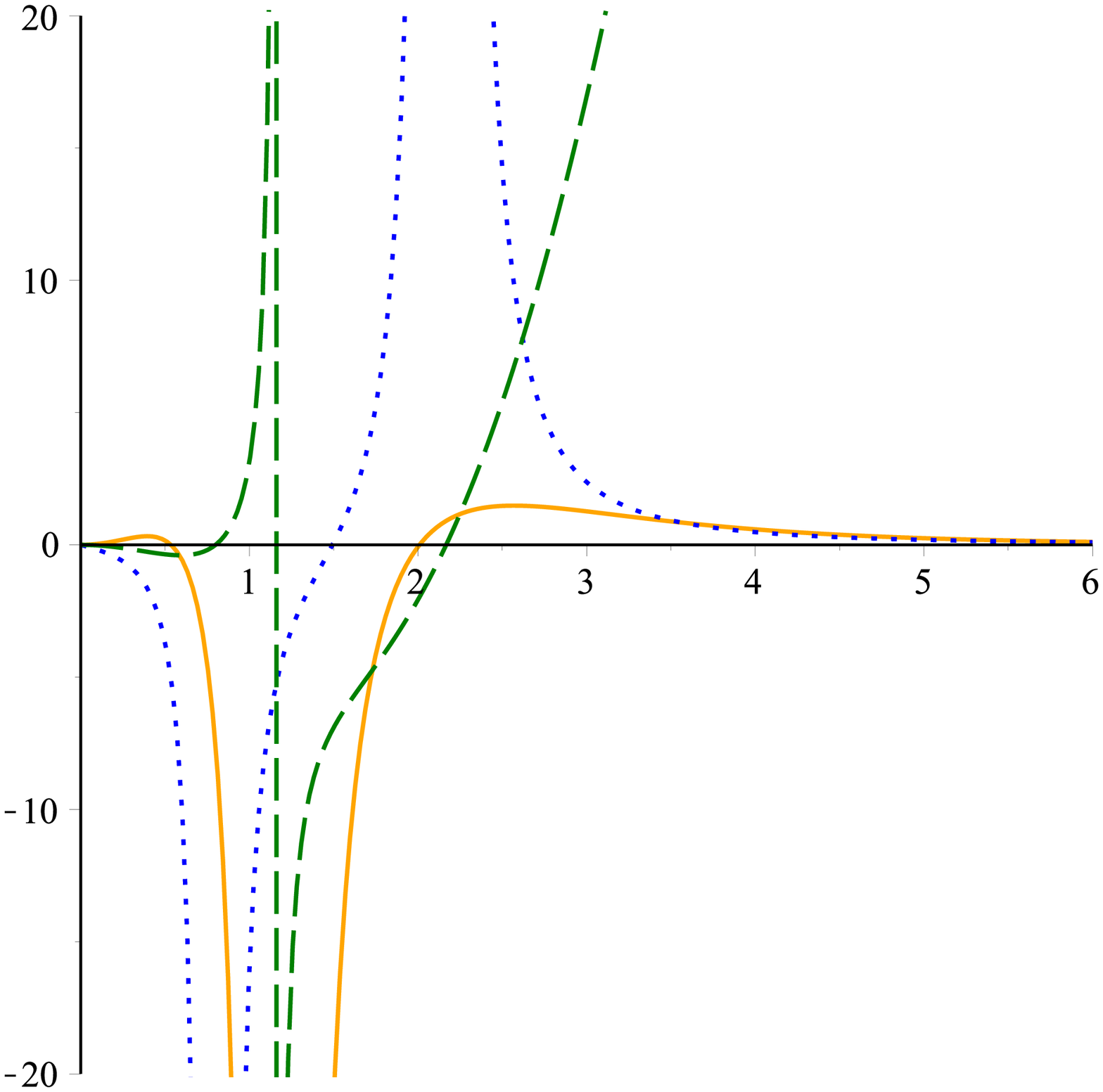}
    }

    \caption{Curvature scalar variation of GTD (orange continuous line),
    Weinhold (blue dot line) metrics, and the heat capacity (green dash line) in terms
    of $ r_{+} $, for $ l=4.0 $ and $ q=0 $, $ q=0.25 $, $ q=0.75 $, for (a), (b) and (c), respectively.}
 \label{pic:CQ1}
 \end{figure}

\begin{figure}[h]
    \centering
     \subfigure[]{
        \includegraphics[width=0.4\textwidth]{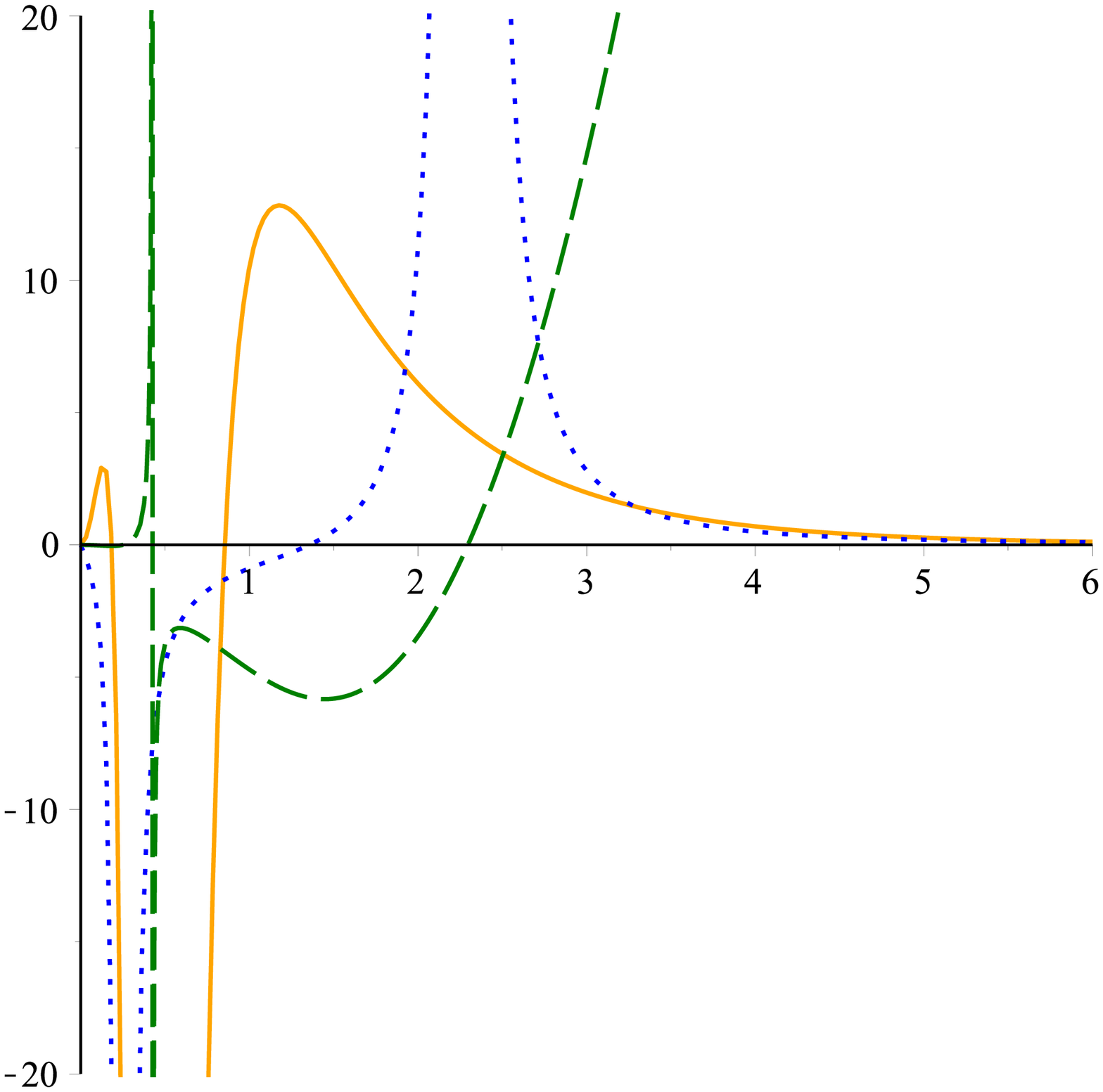}
    }
    \subfigure[]{
        \includegraphics[width=0.4\textwidth]{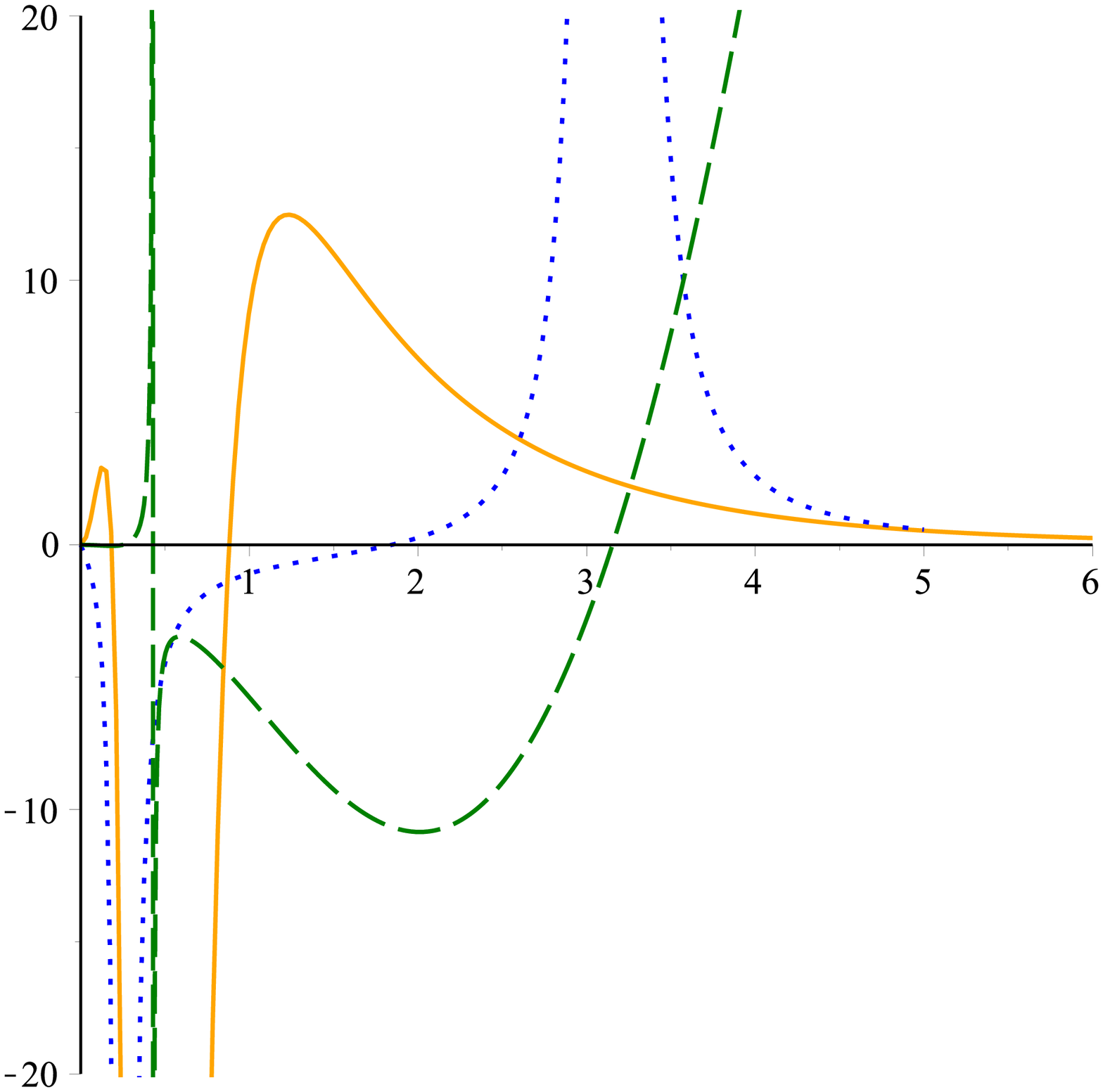}
    }
    \subfigure[]{
        \includegraphics[width=0.4\textwidth]{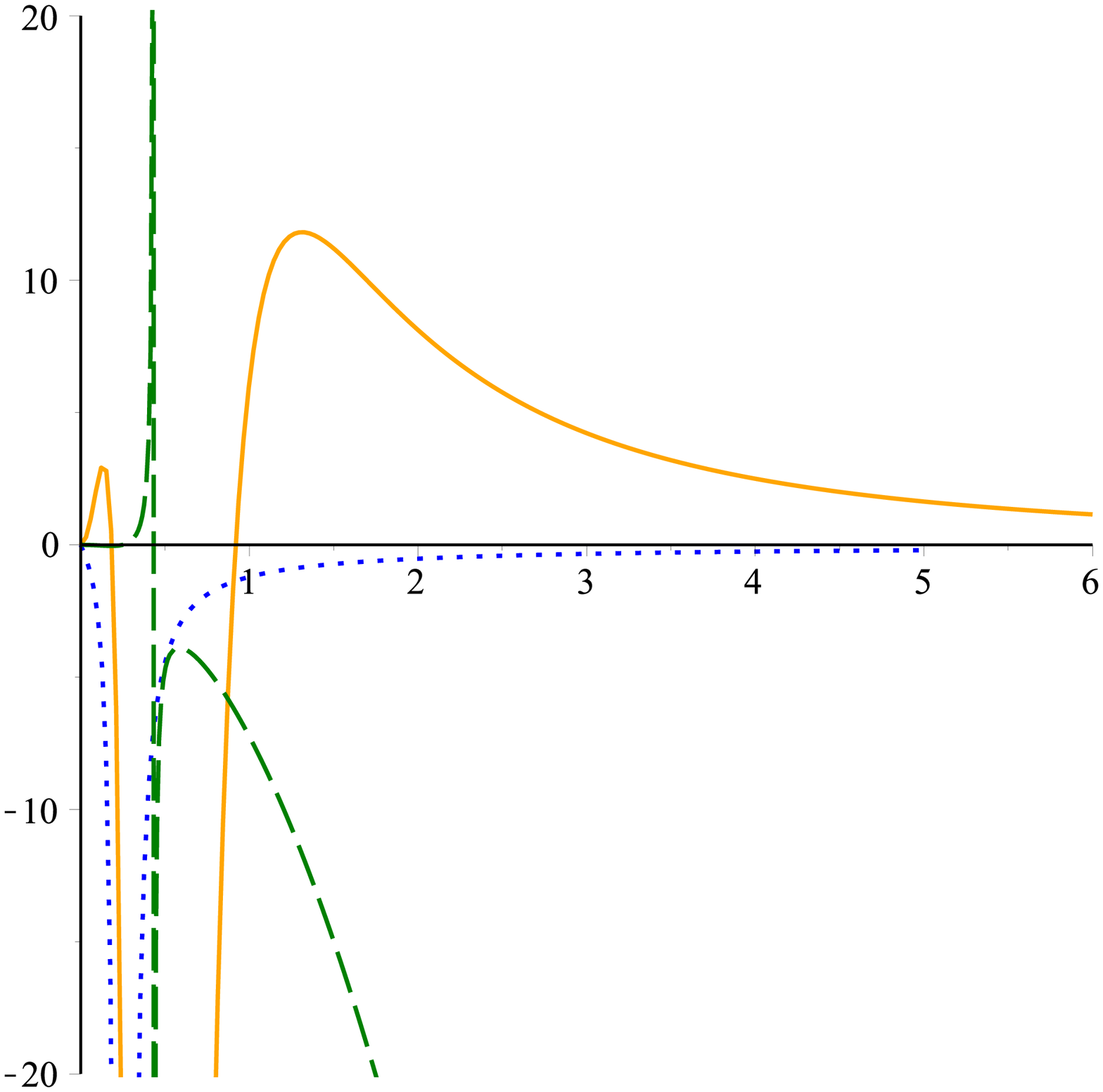}
    }
    \subfigure[]{
        \includegraphics[width=0.4\textwidth]{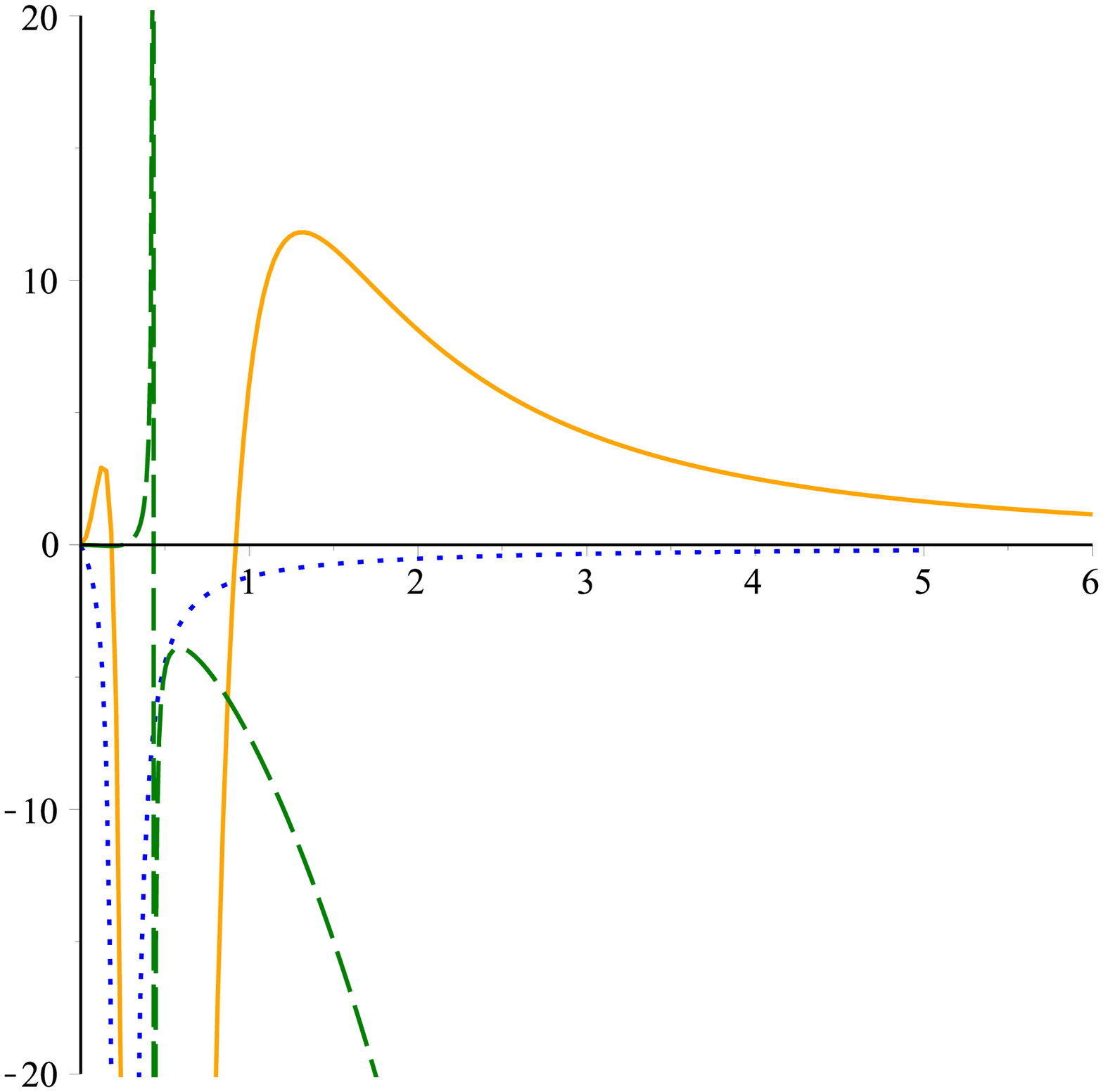}
    }
    \caption{Curvature scalar variation of GTD (orange continuous line), Weinhold (blue dot line) metrics, and the heat capacity (green dash line) in terms of $ r_{+}, $ for $ q=0.25 $ and $ l=4.0 $, $ l=\sqrt{30} $, $ l=\sqrt{3\cdot 10^{4}} $, $ l=\sqrt{3\cdot 10^{15}}$, for (a), (b), (c) and (d), respectively.}
 \label{pic:CQl1}
 \end{figure}

Where, because the numerator of the above expression has no physical
information and it is too long, so, we consider it as $
\mathcal{N} $. The denominator of $ R^{GTD} $, becomes zero at $
r_{+}=r_{\infty} $, (see Fig.~\ref{pic:chargedRGTD}). So, we extended
our study to different thermodynamical geometry. It can be observed
from Fig.~\ref{pic:chargedRGWC}, that, Weinhold and Ruppeiner
methods are compatible with zeros of the heat capacity, and GTD
method is coincide with divergences of it. In following, we point
out the effect of different values of spacetime parameters on stability 
conditions of this black hole. As, it can be seen from Fig.~\ref{pic:CQ1}(a), for
$ q=0 $, the heat capacity of this black hole treats like the black
hole in previous section and It has only one phase transition type
one. By increasing the value of $ q $, it will have two phase
transition type one and one phase transition type two. Also, for
small value of $ l $, heat capacity has two phase transition
type one and one phase transition type two (see,
Fig.~\ref{pic:CQl1}(a,b)), and for large value of $ l $, it has one
phase transition type one and one phase transition type two (see,
Fig.~\ref{pic:CQl1}(c,d)). 

In the next section, we study
thermodynamic behavior of a rotating charged black hole in $ f(R) $
gravity.
\\

\clearpage

 \clearpage

\section{ROTATING CHARGED BLACK HOLE IN $f(R)$ GRAVITY}\label{section4}

In this section, we study solution of the field equation and metric
of a rotating charged black hole in $f(R)$ gravity. With Maxwell
term in four dimensions, the action is
\begin{align}\label{action}
S=S_{g}+S_{M},
\end{align}
where, $S_{g}$, is the gravitational action as

\begin{equation}
S_{g}=\dfrac{1}{16 \pi} \int d^{D} x \sqrt{\vert g \vert }(R+f(R)),
\end{equation}
and $S_{M}$, is the electromagnetic actions as

\begin{equation}\label{action2}
S_{M}=\dfrac{-1}{16 \pi} \int d^{4} x \sqrt{-g }[F_{\mu\nu}F^{\mu\nu}],
\end{equation}
where, $R$, is the scalar curvature and $R + f(R)$, is the function
defining the theory under consideration, and $g$, is the determinant
of the metric. From the Eq.~(\ref{action2}), the Maxwell equation
takes the form
\begin{equation}
\nabla_{\mu}F^{\mu\nu}=0.
\end{equation}
The field equations in the metric formalism are~\cite{Larranaga:2011fv}
\begin{align}\label{rmiyo}
R_{\mu\nu}\big(1+ f'(R)\big)-\frac{1}{2}\big(R+f(R)\big)g_{\mu\nu}
+\big(g_{\mu\nu}\nabla^{2}-\nabla_{\mu}\nabla_{\nu}\big)f'(R)=2T_{\mu\nu},
\end{align}
where  $\nabla$ is the usual covariant derivative, $R_{\mu\nu}$, is
the Ricci tensor and $ T_{\mu\nu} $, the stress-energy tensor of the
electromagnetic field as
\begin{equation}
T_{\mu\nu}=F_{\mu\rho}F_{\nu}^{\rho}-\dfrac{g_{\mu\nu}}{4}F_{\rho\sigma}F^{\rho\sigma},
\end{equation}
with
\begin{equation}
T^{\mu} _{\,\,\mu}=0 .
\end{equation}
The trace of Eq.~(\ref{rmiyo}) with the constant curvature scalar
$R=R_{0}$, yields
\begin{equation}
R_{0}\big(1+f'(R_{0})\big)-2\big(R_{0}+f(R_{0})\big)=0,
\end{equation}
which determines the negative constant curvature scalar as
\begin{equation}\label{R01}
R_{0}=\dfrac{2f(R_{0})}{f'(R_{0})-1} .
\end{equation}
Using Eqs.~(\ref{rmiyo})--(\ref{R01}), we have
\begin{align}
R_{\mu\nu}=\dfrac{1}{2}\big(\dfrac{f(R_{0})}{f'(R_{0})-1}\big)g_{\mu\nu}
+\dfrac{2}{\big(1+f'(R_{0})\big)}T_{\mu\nu}.
\end{align}
Finally, the axisymmetric ansatz in Boyer--Lindquist--type
coordinates $(t,r,\theta,\varphi)$, inspired by the Kerr-Newman-AdS black hole solution, is~\cite{Larranaga:2011fv}
\begin{align}
ds^{2}=-\dfrac{\Delta_{r}}{\rho^{2}} \big[dt-\dfrac{a sin^{2}\theta
d\varphi}{\Xi}\big]^{2}
+\dfrac{\rho^{2}}{\Delta_{r}}dr^{2}+\dfrac{\rho^{2}}{\Delta_{\theta}}d\theta^{2}
+\dfrac{\Delta_{\theta}sin^{2}\theta}{\rho^{2}}\big[adt-\dfrac{r^{2}+a^{2}}{\Xi}d\varphi
\big]^{2},
\end{align}
where
\begin{align}
\Delta_{r}=(r^{2}+a^{2}) \big(1+\dfrac{R_{0}}{12}r^{2} \big)-2mr+\dfrac{Q^{2}}{(1+f'(R_{0}))},
\end{align}
\begin{align}
\Xi=1-\dfrac{R_{0}}{12}a^{2}, \qquad
\rho^{2}=r^{2}+a^{2}cos^{2}\theta, \qquad
\Delta_{\theta}=1-\dfrac{R_{0}}{12}a^{2}cos^{2}\theta ,
\end{align}
in which $ R_{0} $, is a constant proper to cosmological constant ($R_{0}=-4\Lambda $), 
$ Q $, 
is the electric charge and $ a $, is the
angular momentum per mass of the black hole.
\\

\textbf{4.1.}  \textbf{Thermodynamic}

In this section, we investigate the thermodynamic peroperties of this black hole. The radius of the horizon $ (r_{+}) $ satisfy the condition $ \Delta_{r}=0 $,

\begin{equation}\label{Delta}
(r^{2}_{+}+a^{2})\left( 1+\frac{R_{0}}{12}r^{2}_{+}\right) -2mr_{+}+\frac{Q^{2}}{(1+f^{\prime}(R_{0}))}=0.
\end{equation}

By setting $ dr=dt=0 $, in the metric line elements, we can find
line elements for the 2-Dimensional horizon. Using the relation

\begin{equation}
A=\int^{2\pi}_{0}d\varphi\int^{\pi}_{0}\sqrt{|\gamma |}d\theta,
\end{equation}
where $ \gamma $, is the metric tensor of the black hole horizon, the area of this black hole will be obtained as

\begin{equation}\label{area}
A=\frac{4\pi(r^{2}_{+}+a^{2})}{1-\frac{R_{0}}{12}a^{2}}.
\end{equation}

According to relation for entropy, $ S=\frac{A}{4} $~\cite{Bekenstein:1973ur}, we can easily find the entropy of this black hole as,

\begin{equation}\label{entropy}
S=\frac{\pi(r^{2}_{+}+a^{2})}{1-\frac{R_{0}}{12}a^{2}}.
\end{equation}

Mass of the black hole can be obtained by using generalized Smarr
formula in terms of all its parameter. To calculate generalized
Smarr formula, first we obtain total mass $ (M) $, and angular
momentum $ (J) $, by means of Kommar integrals and using the killing
vectors, $ \frac{1}{\Xi}\partial_{t} $, and $ \partial_{\varphi} $,
so they will be obtained as

\begin{equation}
M=\frac{m}{\Xi^{2}},
\end{equation}

\begin{equation}\label{J}
J=\frac{am}{\Xi^{2}}.
\end{equation}

Using Eqs.~(\ref{Delta})--(\ref{J}), the generalized Smarr formula
will be obtained as

\begin{equation}
M^{2}=\frac{S}{4\pi}+\frac{\pi}{4S}\left[4J^{2}+q^{4}\right]+\frac{q^{2}}{2}-\frac{R_{0}}{12}J^{2}-\frac{R_{0}S}{24\pi}\left[q^{2}+\frac{S}{\pi}-\frac{R_{0}S^{2}}{24\pi^{2}}\right].
\end{equation}
Now, according to the first law of thermodynamic, we can calculate
all of the thermodynamic quantities,

\begin{equation}
dM=TdS+\Omega dJ+\Phi dq.
\end{equation}
So, the temperature of this black hole is,
\begin{align}\label{T}
T=\frac{\partial M}{\partial S}=-(48q^{4}\pi^{5}+8R_{0}\pi^{3}R_{0}S^{2}q^{2}+192\pi^{5}J^{2}-R_{0}^{2}\pi S^{4}+16R_{0}\pi^{2}S^{3}-48\pi^{3}S^{2})\cdot \nonumber\\
(256\pi^{4}S^{3}(144\pi^{5}q^{4}-48R_{0}\pi^{4}J^{2}S-24R_{0}\pi^{3}q^{2}S^{2}+576\pi^{5}J^{2}+\nonumber\\
288\pi^{4}q^{2}S+R_{0}^{2}\pi S^{4}-24R_{0}\pi^{2}S^{3}+144\pi^{3}S^{2})^{-\frac{1}{2}},
\end{align}
In addition, the angular velocity $ \Omega $ is,
\begin{align}
\Omega =\frac{\partial M}{\partial J}=-(2\pi^{2} J(R_{0}S-12\pi))\cdot(S(144\pi^{5}q^{4}-48R_{0}\pi^{4}J^{2}S-24R_{0}\pi^{3}q^{2}S^{2}+576\pi^{5}J^{2}+\nonumber\\
288\pi^{4}q^{2}S+R_{0}^{2}\pi S^{4}-24R_{0}\pi^{2}S^{3}+144\pi^{3}S^{2}))^{-\frac{1}{2}},
\end{align}
and, also the electrical potential can be obtained as,
\begin{align}
\Phi =\frac{\partial M}{\partial q}=-\pi q(-12\pi^{2}q^{2}+R_{0}S^{2}-12\pi S)\cdot (S(144\pi^{5}q^{4}-48R_{0}\pi^{4}J^{2}S-24R_{0}\pi^{3}q^{2}S^{2}+\nonumber\\
576\pi^{5}J^{2}+288\pi^{4}q^{2}S+
R_{0}^{2}\pi S^{4}-24R_{0}\pi^{2}S^{3}+144\pi^{3}S^{2}))^{-\frac{1}{2}}.
\end{align}

Finally, we can calculate the heat capacity of this black hole as follows

\begin{equation}
C=\frac{\partial_{S}M}{\partial^{2}_{S}M}.
\end{equation}

\begin{figure}[h]
    \centering
        \includegraphics[width=0.4\textwidth]{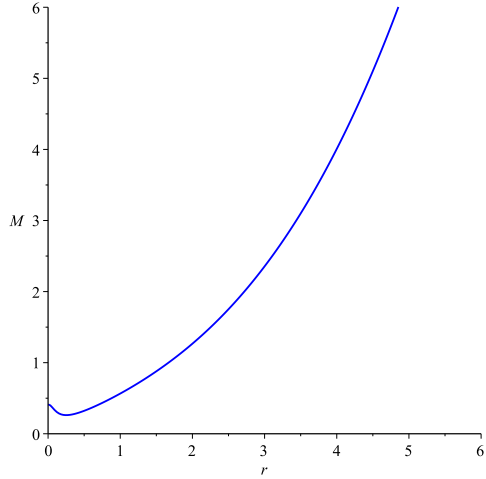}
    \caption{Mass variation of a Rotating charged black hole in terms of its horizon radius $ r_{+} $ for $ l=4.0 $, $ a=0.1 $, $ q=0.25 $.}
 \label{pic:rotatingM}
\end{figure}

\begin{figure}[h]
    \centering
        \includegraphics[width=0.4\textwidth]{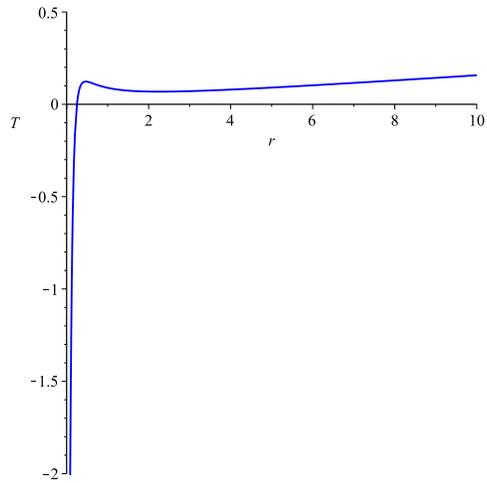}
    \caption{Temperature variation of a Rotating charged black hole in terms of its horizon radius $ r_{+} $ for $ l=4.0 $, $ a=0.1 $, $ q=0.25 $.}
 \label{pic:rotatingT}
\end{figure}

\begin{figure}[h]
    \centering
     \subfigure[]{
        \includegraphics[width=0.4\textwidth]{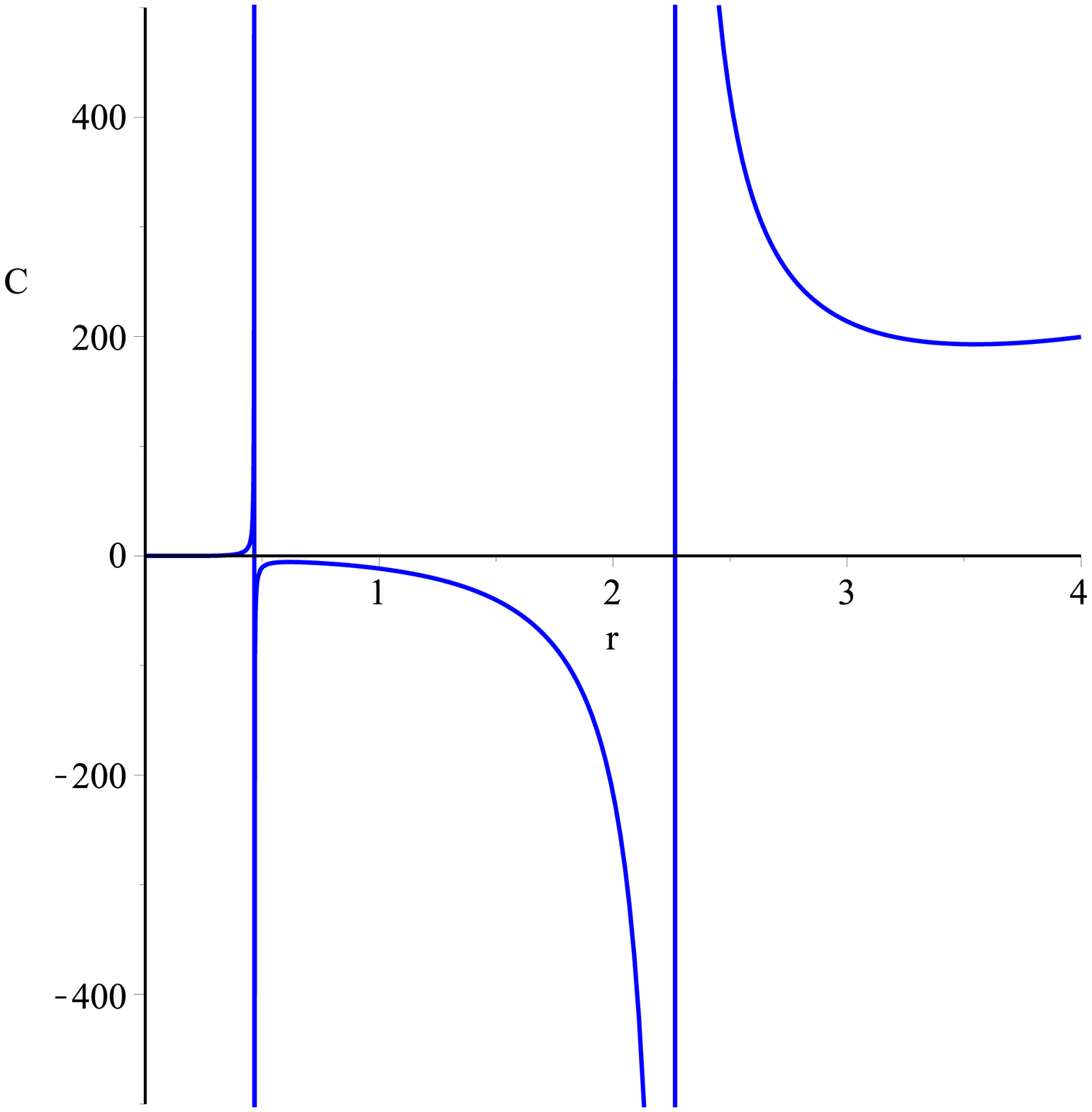}
    }
    \subfigure[Closeup of figure (a)]{
        \includegraphics[width=0.4\textwidth]{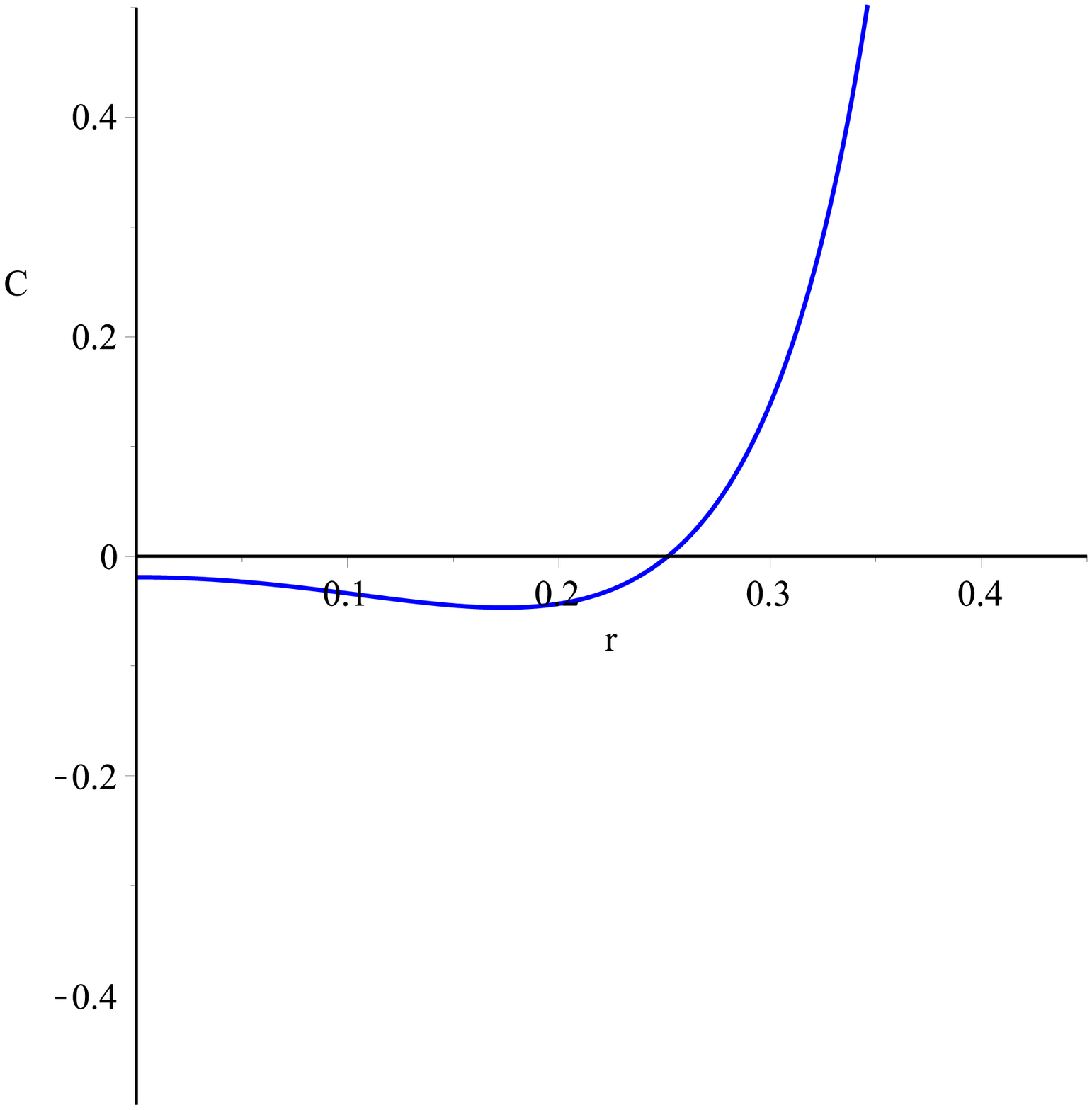}
    }
    \caption{Heat capacity variation of a Rotating charged black hole in terms of horizon radius $ r_{+} $, for $ l=4.0 $, $ a=0.1 $, $ q=0.25 $.}
 \label{pic:rotatingC}
\end{figure}

Plot of all thermodynamic parameters obtained for this black hole, are shown in Figs.~\ref{pic:rotatingM}-~\ref{pic:rotatingC}.

\clearpage

It can be seen from Fig.~\ref{pic:rotatingM}, the mass of this black
hole, has one minimum point at $ r_{+}=r_{m} $, (we show the place
of minimum point of the mass with $ r_{m} $), in which the value of
it, is equal to 0.252. It is also observed from the plot of
temperature in Fig.~\ref{pic:rotatingT}, that, the temperature of
this system is in the negative region at a particular range of $
r_{+} $ ($ r_{+}<r_{m} $), after that, it reaches to zero at $
r_{+}=r_{m} $, then, it will be positive for $ r_{+}>r_{m} $. In
addition, Fig.~\ref{pic:rotatingC}, shows that, the heat capacity of
this black hole arrives to positive (stable) phase from negative
(unstable) phase, after that, it reaches to zero at $ r_{+}=r_{m} $.
Also the divergence points of heat capacity are $ r_{\infty 1} $ and
$ r_{\infty 2} $, that, for this system $ r_{\infty 1}=0.466 $ and $
r_{\infty 2}=2.266 $. So, for the range of $ r_{m}<r_{+}<r_{\infty
1} $, the heat capacity is positive and system is in the stable
phase, after that, at $ r_{\infty 1}<r_{+}<r_{\infty 2} $, it falls
in to negative region (unstable phase), then at $ r_{+}>r_{\infty 2}
$ it will be positive(stable). In other words, the heat capacity of
this black hole, has one phase transition type one, and two phase
transition type two.
\\

\textbf{4.2.}  \textbf{Thermodynamic geometry}

In this part, we investigate thermodynamic geometry of this black
hole, using Weinhold, Ruppeiner and GTD methods. We start by
Weinhold metric, which is as follows

\begin{equation}
g^{W}=\begin{bmatrix}
M_{SS} & M_{SJ} & M_{Sq} & M_{Sl}\\
M_{JS} & M_{JJ} & M_{Jq} & M_{Jl}\\
M_{qS} & M_{qJ} & M_{qq} & M_{l q}\\
M_{l S} & M_{l J} & M_{l q} & M_{ll}
\end{bmatrix}.
\end{equation}

\begin{figure}[h]
    \centering
        \includegraphics[width=0.4\textwidth]{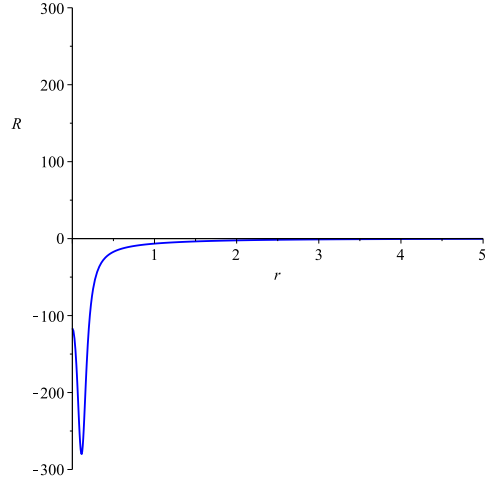}
    \caption{Curvature scalar variation of Weinhold metric in terms of horizon radius $ r_{+} $ for $ l=4.0 $, $ q=0.25 $, $ a=0.1 $.}
 \label{pic:rotatingRW}
\end{figure}

The scalar curvature $ R^{W} $, can be easily obtained, which is
plotted in Fig.~\ref{pic:rotatingRW}.

It can be seen from Fig.~\ref{pic:rotatingRW}, The curvature scalar has no singularity, 
so Weinhold method has no physical information for this system.

\begin{figure}[h]
    \centering
        \includegraphics[width=0.4\textwidth]{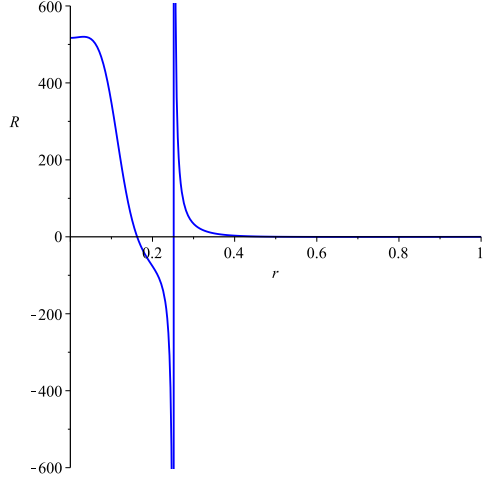}
    \caption{Curvature scalar variation of Ruppeiner metric in terms of horizon radius $ r_{+} $. $ l=4.0 $ for $ q=0.25 $, $ a=0.1 $.}
 \label{pic:rotatingRR}
\end{figure}

Now, we construct Ruppeiner metric for this black hole as follows

\begin{equation}
g^{R}=\frac{1}{T}\begin{bmatrix}
M_{SS} & M_{SJ} & M_{Sq} & M_{Sl}\\
M_{JS} & M_{JJ} & M_{Jq} & M_{Jl}\\
M_{qS} & M_{qJ} & M_{qq} & M_{l q}\\
M_{l S} & M_{lJ} & M_{l q} & M_{ll}
\end{bmatrix},
\end{equation}
where, $ T $, can be obtained from Eq.~(\ref{T}).
The curvature scalar, which is correspond to above metric, is plotted in Fig.~\ref{pic:rotatingRR}, and it is singular at $ r_{+}=r_{m} $.

Finally, at the end of this section, we apply the most important metric of GTD method to this thermodynamic system, as

\begin{equation}
g^{GTD}=\begin{bmatrix}
-M_{SS} & 0 & 0 & 0\\
0 & M_{JJ} & M_{Jq} & M_{Jl}\\
0 & M_{qJ} & M_{qq} & M_{l q}\\
0 & M_{l J} & M_{l q} & M_{ll}
\end{bmatrix}.
\end{equation}

\begin{figure}[h]
    \centering
        \includegraphics[width=0.4\textwidth]{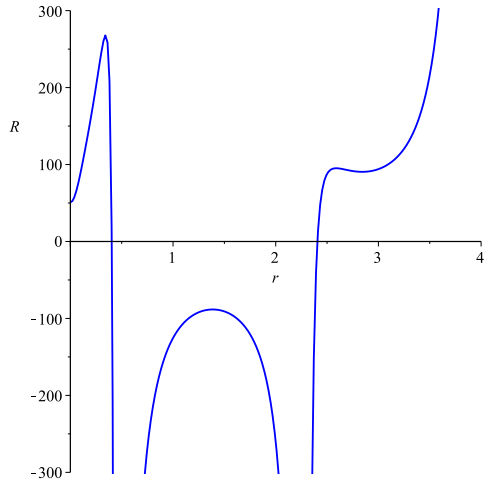}
    \caption{Curvature scalar variation of GTD metric in terms of horizon radius $ r_{+} $ for $ l=4.0 $, $ a=0.1 $ and $ q=0.25$.}
 \label{pic:rotatingRGTD}
\end{figure}

 Plot of the corresponding curvature scalar with GTD metric, is shown in Fig.~\ref{pic:rotatingRGTD}.\\

\begin{figure}[h]
    \centering
        \includegraphics[width=0.4\textwidth]{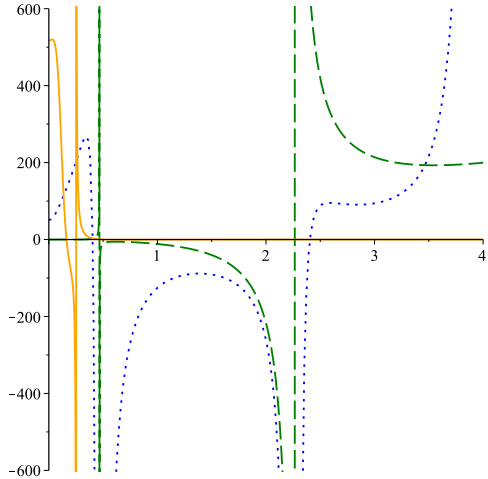}
    \caption{Curvature scalar variation of Ruppeiner (orange continuous line), GTD (blue dot line) metrics, and the heat capacity (green dash line) of a rotating black hole in terms of $ r_{+} $ for $ l=4.0 $, $ a=0.1 $, $ q=0.25 $.}
 \label{pic:rotatingRGC}
\end{figure}

This curvature scalar is singular at $ r_{+}=r_{\infty 1} $ and $
r_{+}=r_{\infty 2} $. So, again we extended our study to different
geothermodynamic methods, and our results are shown in
Fig.~\ref{pic:rotatingRGC}. It can be observed from
Fig.~\ref{pic:rotatingRGC}, singularities of Ruppeiner metric is
compatible with the zero point of heat capacity, and singularities
of GTD metric, are coincide with divergence points of heat capacity.
At the end of this section, we investigate the effect of different
values of $ q $, $ a $, and $ l $, parameters on phase transition
points for this system.
\\

\begin{figure}[h]
    \centering
     \subfigure[]{
        \includegraphics[width=0.4\textwidth]{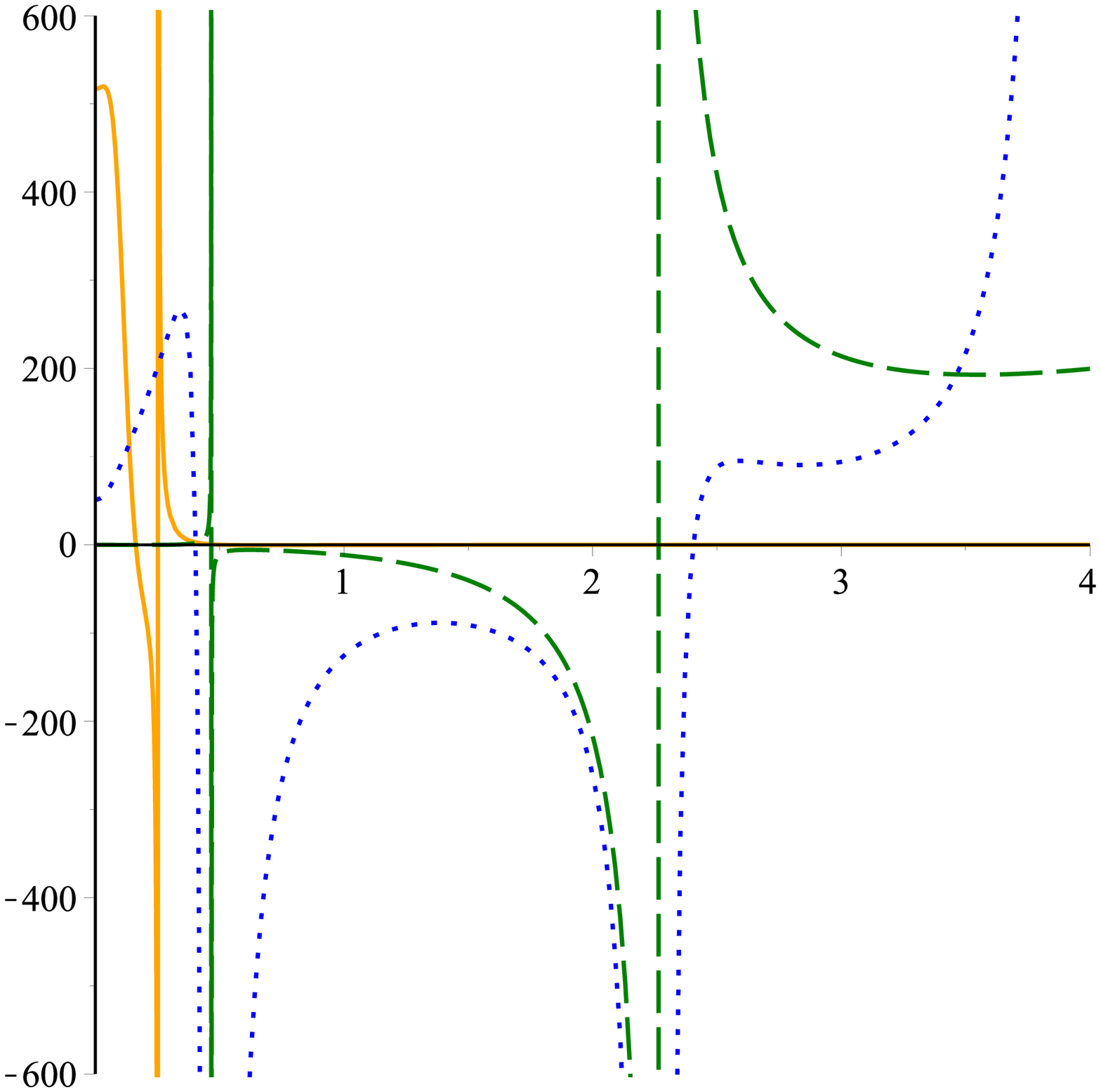}
    }
    \subfigure[]{
        \includegraphics[width=0.4\textwidth]{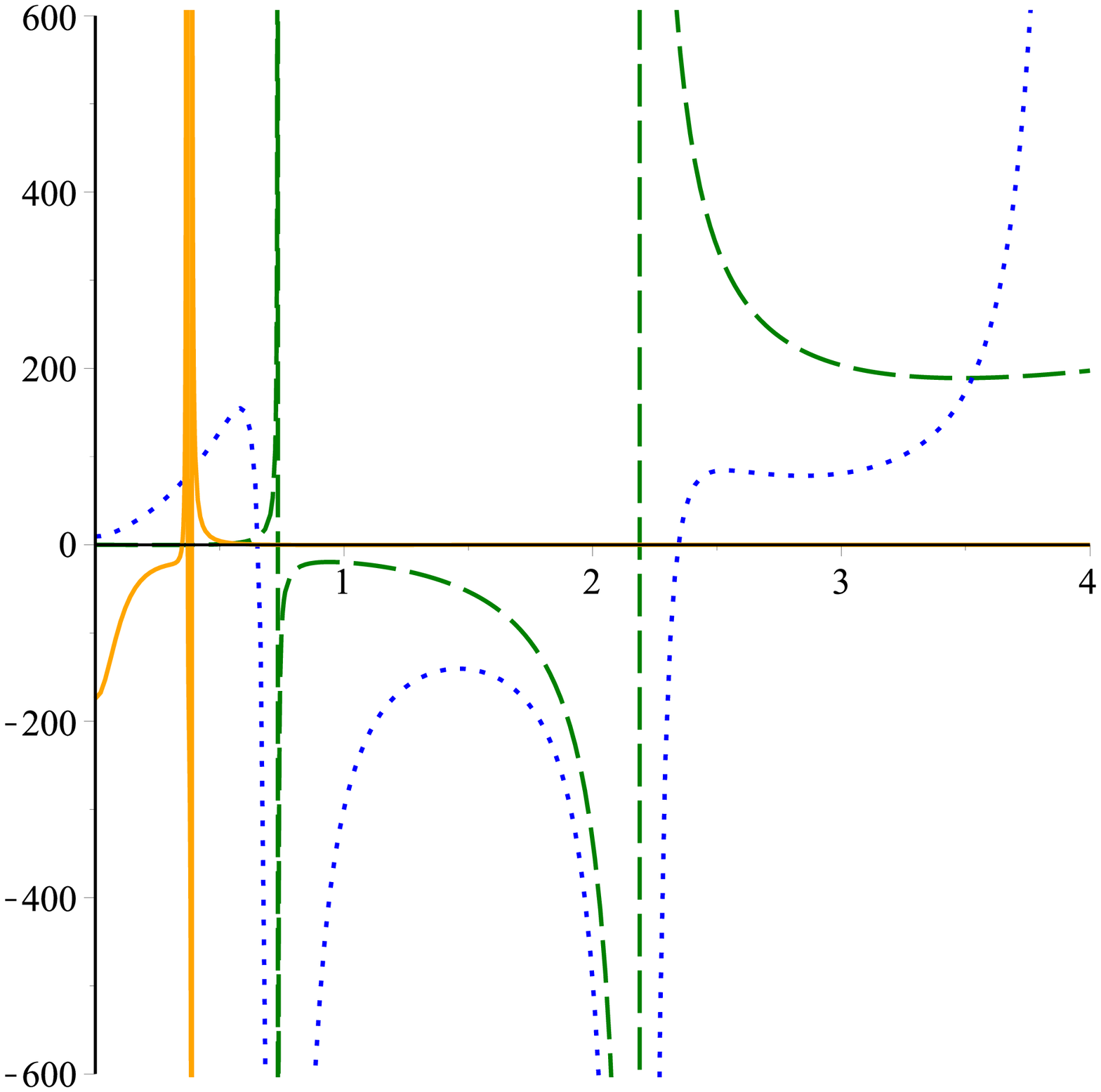}
    }
    \subfigure[]{
        \includegraphics[width=0.4\textwidth]{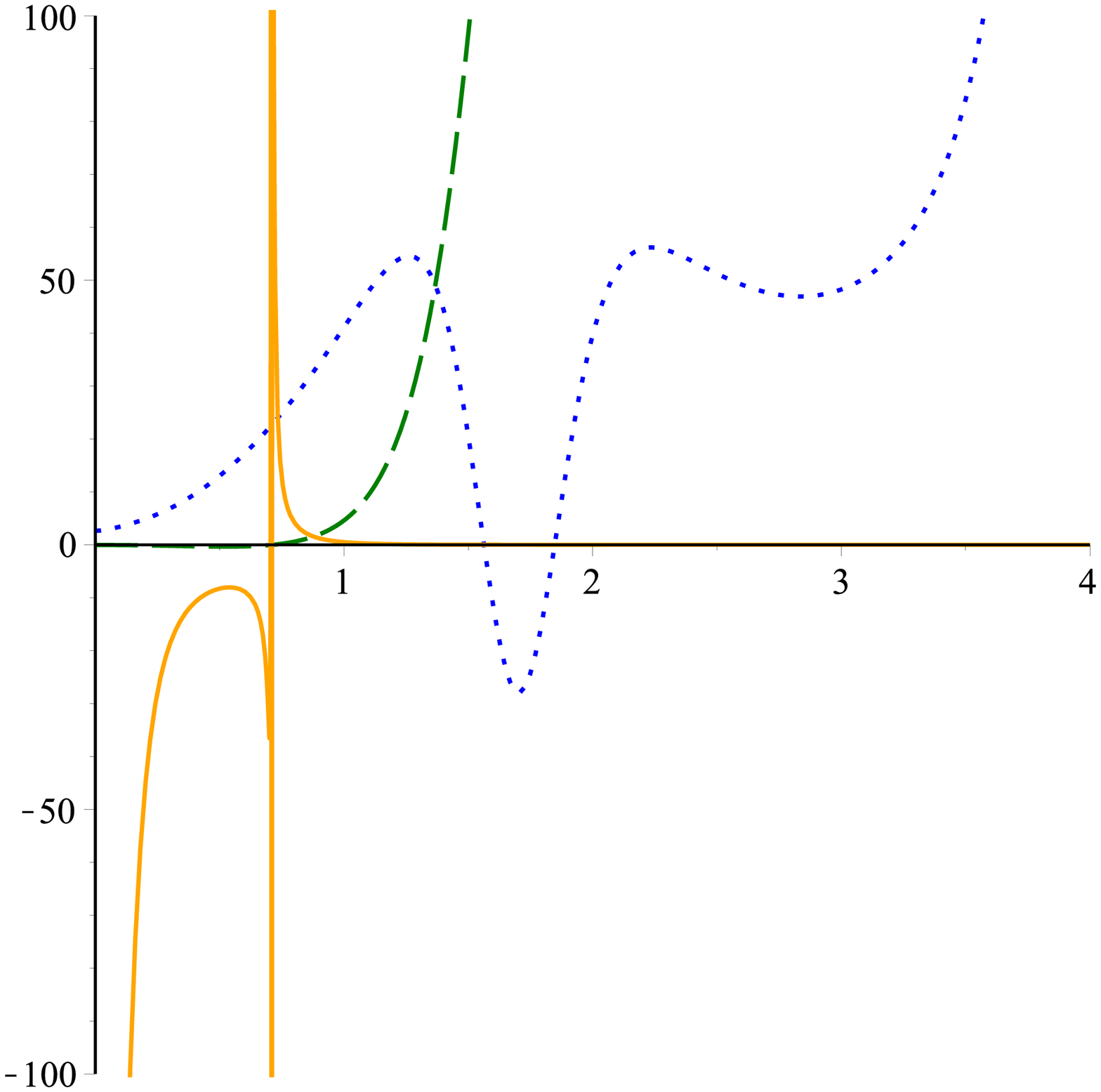}
    }
    \subfigure[]{
        \includegraphics[width=0.4\textwidth]{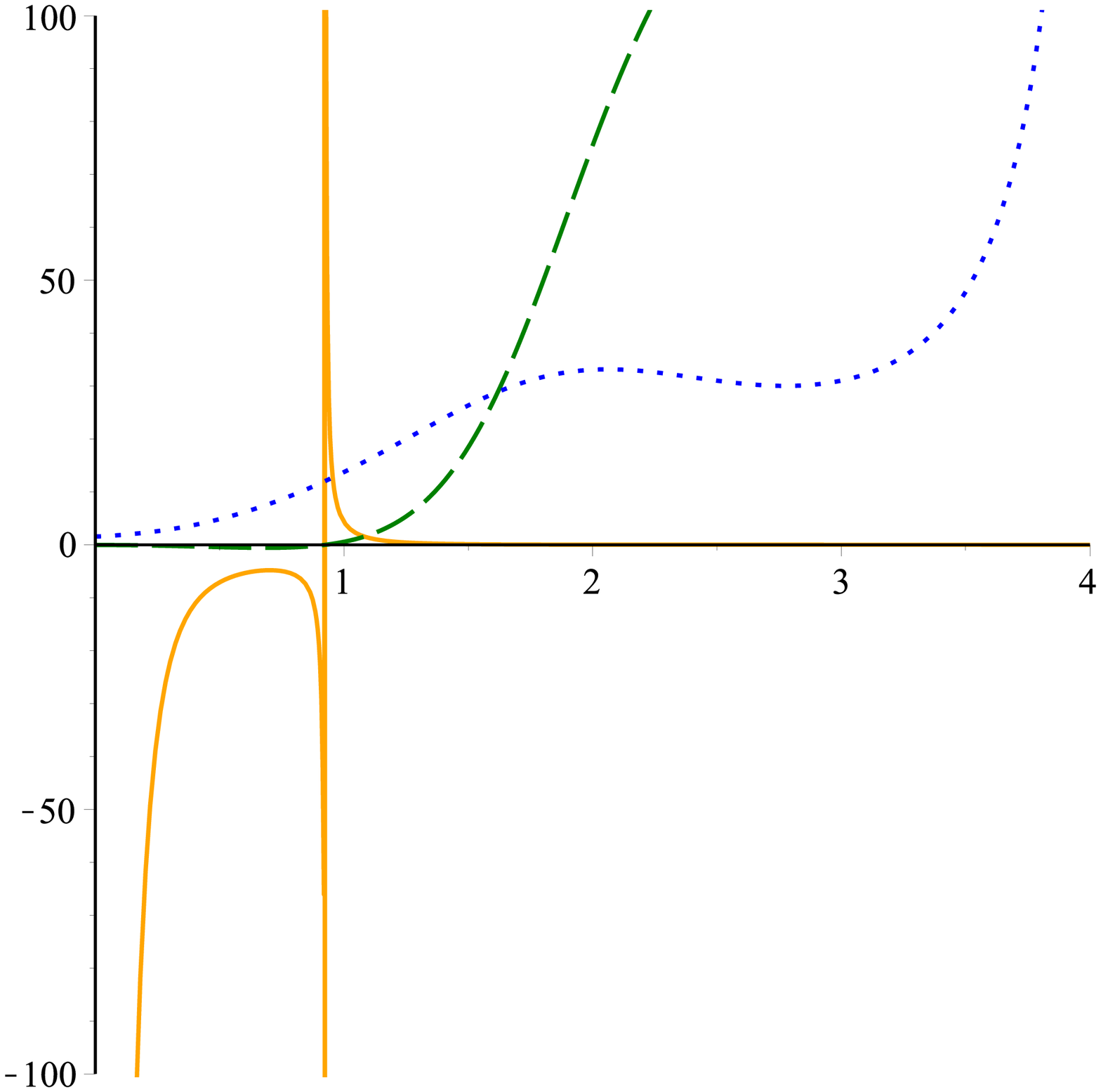}
    }
    \caption{Curvature scalar variation of Ruppeiner (orange continuous line), GTD (blue dot line) metrics and the heat capacity (green dash line) of a rotating black hole in terms of $ r_{+} $ for $ l=4.0 $, $ a=0.1 $ and $ q=0.25 $, $ q=0.4 $, $ q=0.75 $, $ q=1 $, for (a), (b), (c) and (d), respectively.}
 \label{pic:CRq}
 \end{figure}

\begin{figure}[h]
    \centering
     \subfigure[]{
        \includegraphics[width=0.4\textwidth]{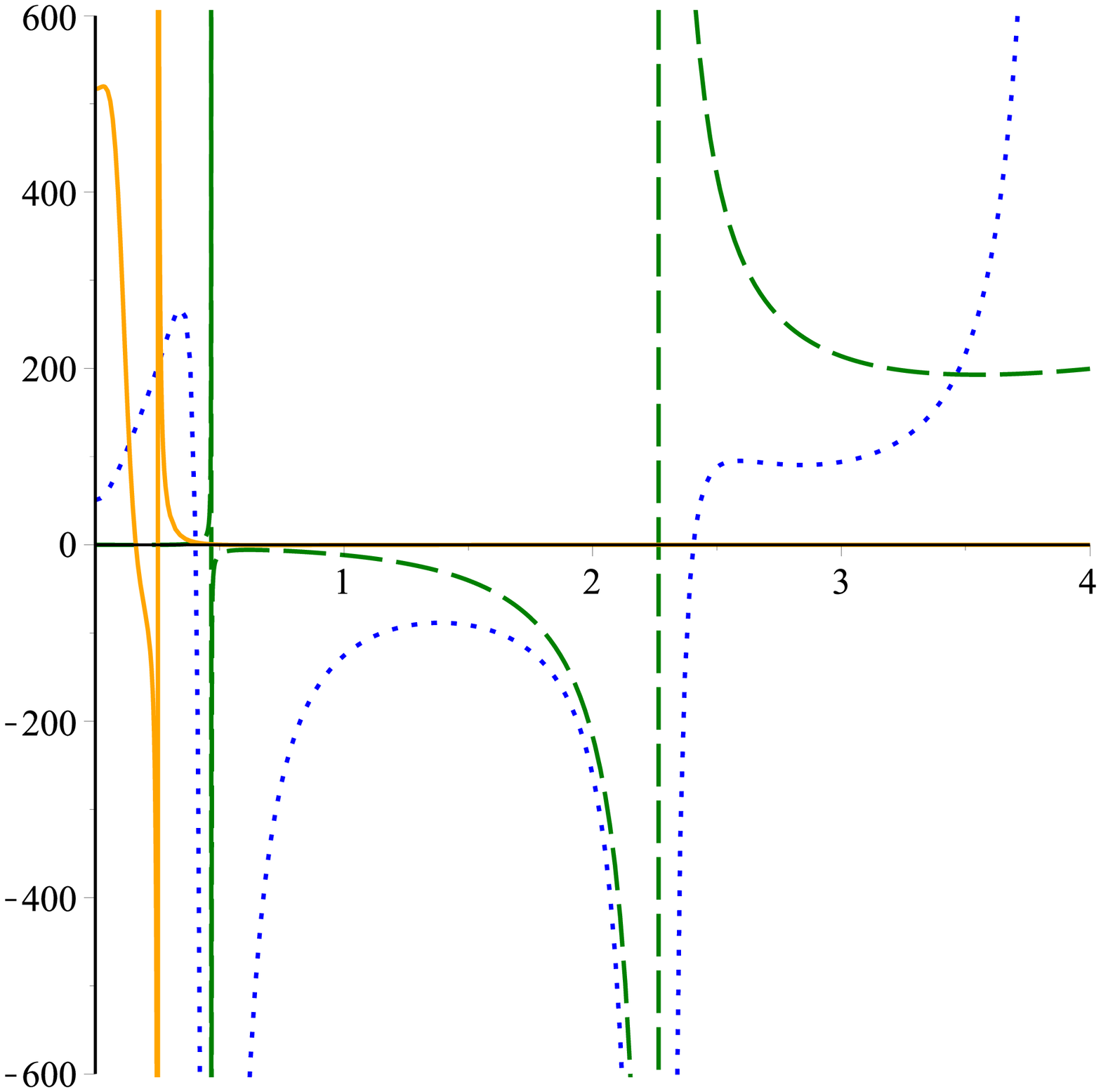}
    }
    \subfigure[]{
        \includegraphics[width=0.4\textwidth]{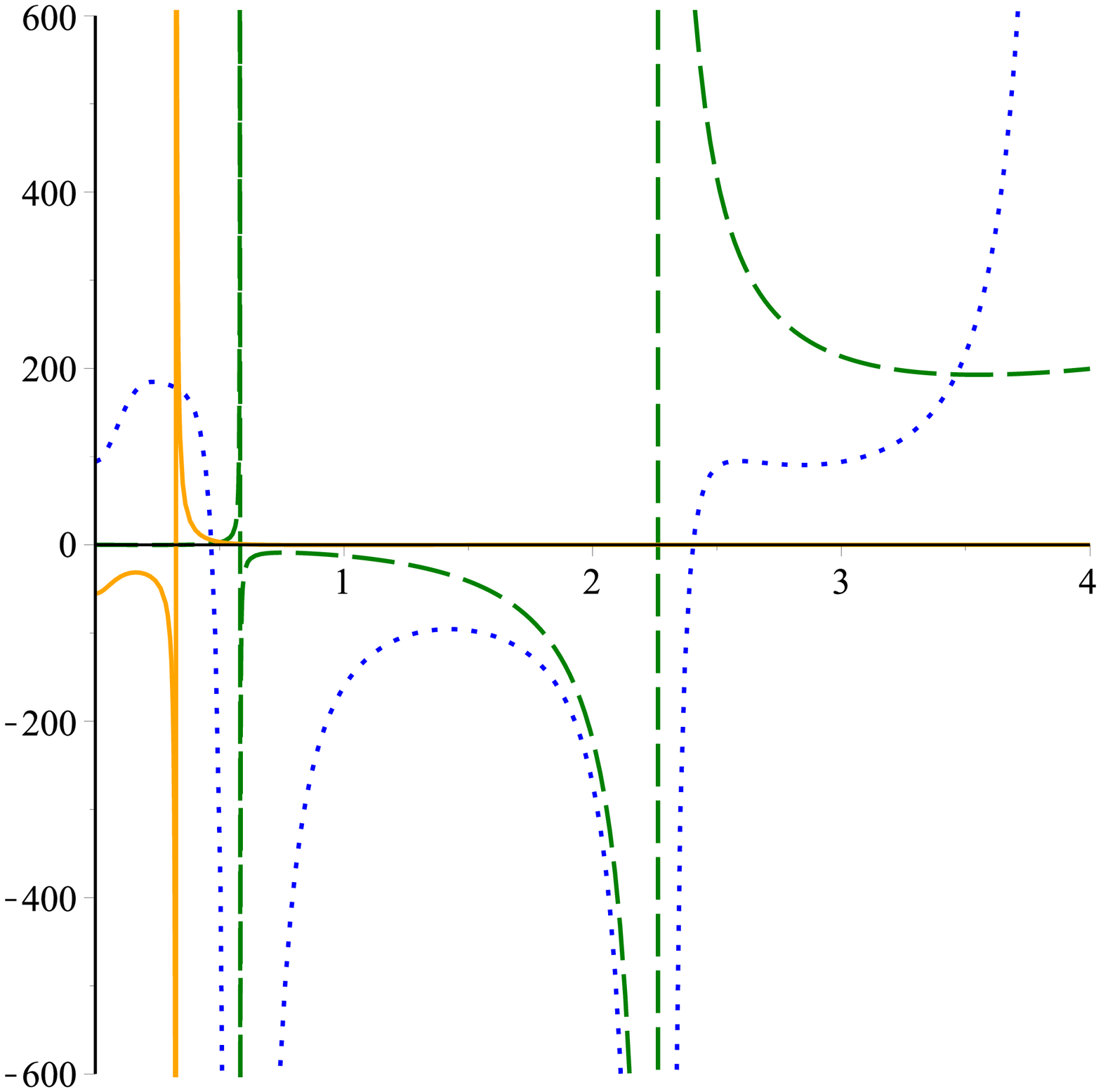}
    }
    \subfigure[]{
        \includegraphics[width=0.4\textwidth]{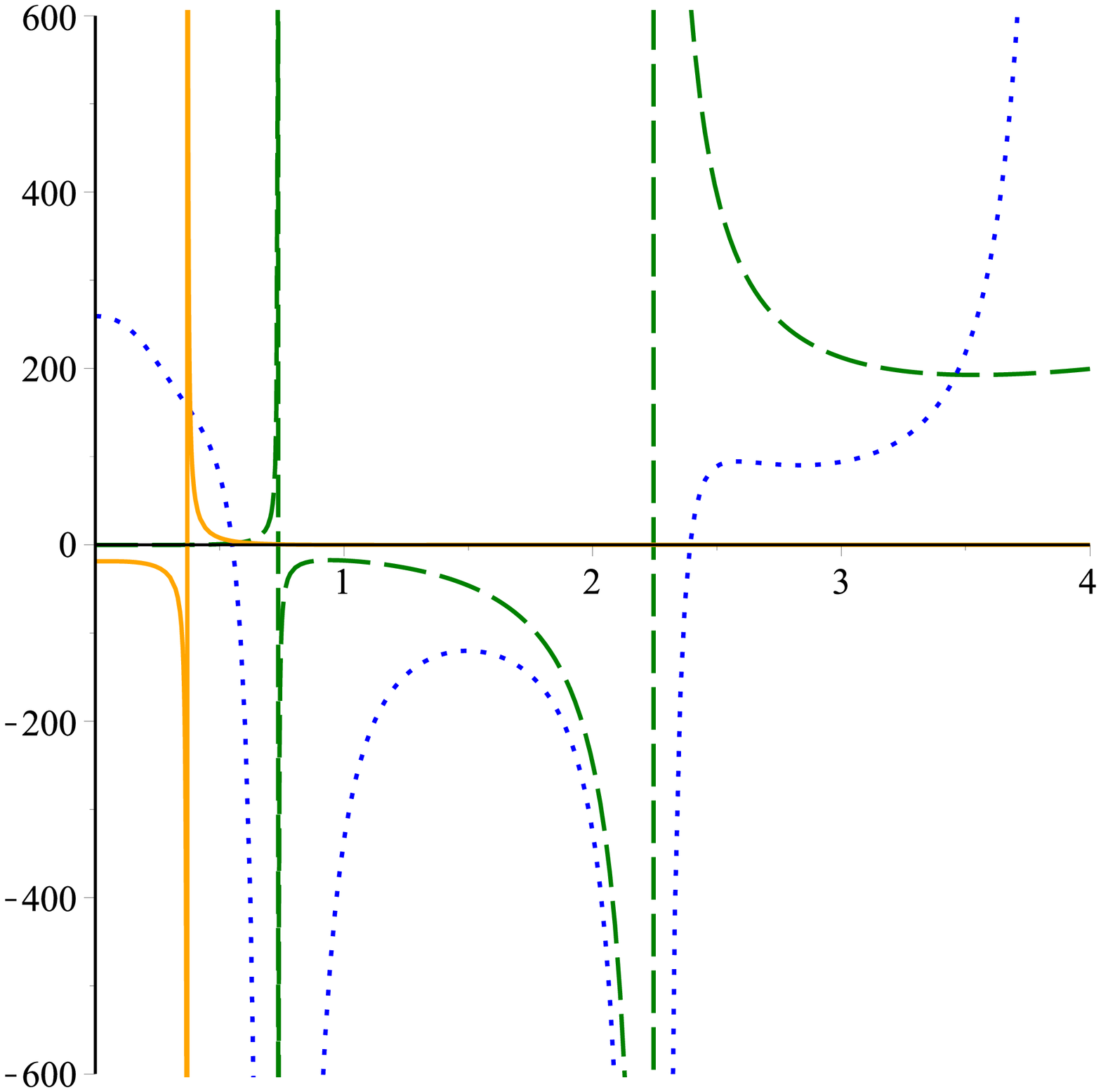}
    }
    \subfigure[]{
        \includegraphics[width=0.4\textwidth]{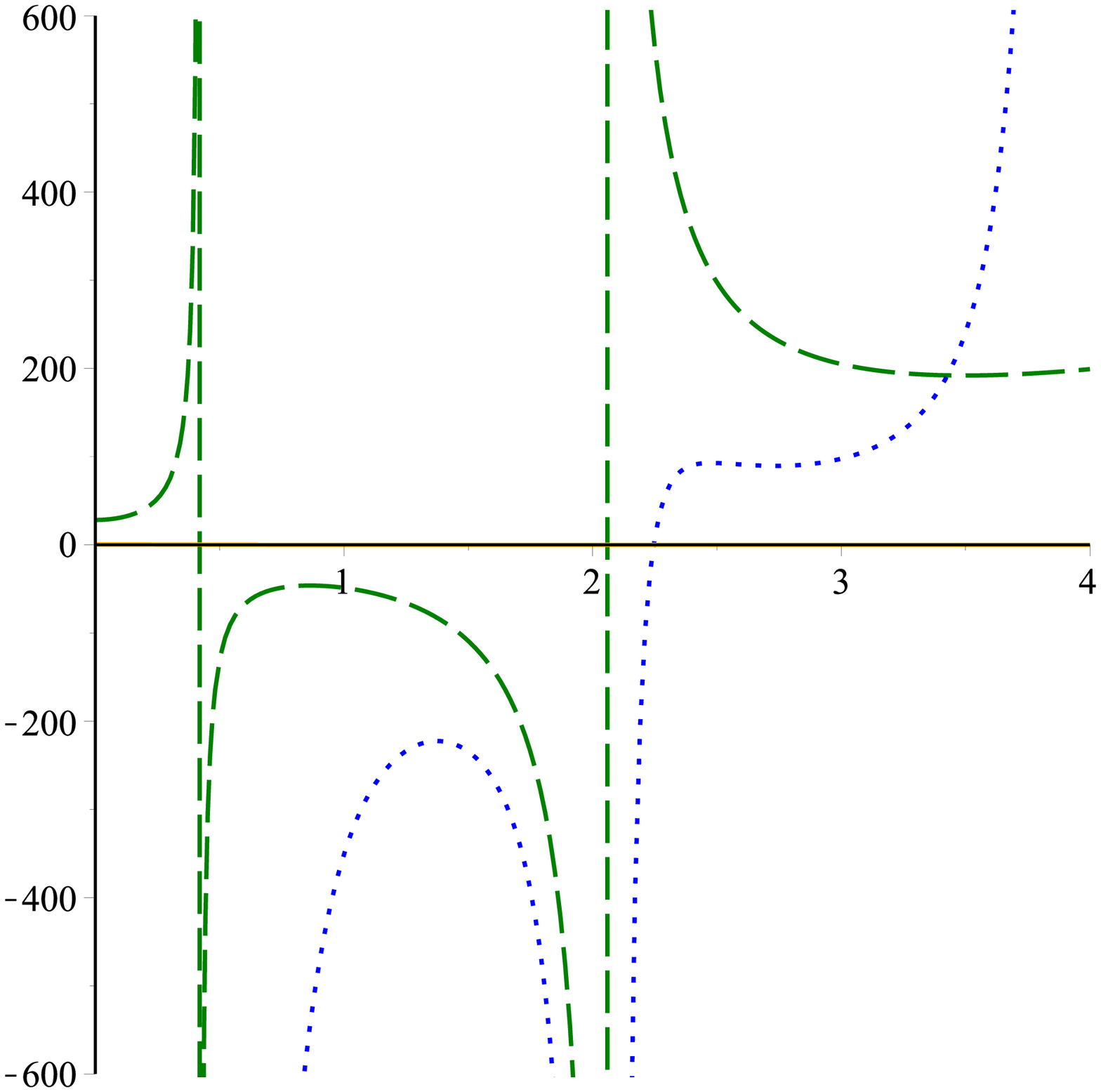}
    }
    \caption{Curvature scalar variation of Ruppeiner (orange continuous line), GTD (blue dot line) metrics, and the heat capacity (green dash line) of a rotating black hole in terms of $ r_{+} $ for $ l=4.0 $, $ q=0.25 $ and $ a=0.1 $, $ a=0.25 $, $ a=0.5 $, $ a=0.8 $, for (a), (b), (c) and (d), respectively.}
 \label{pic:CRa}
 \end{figure}

\begin{figure}[h]
    \centering
     \subfigure[]{
        \includegraphics[width=0.4\textwidth]{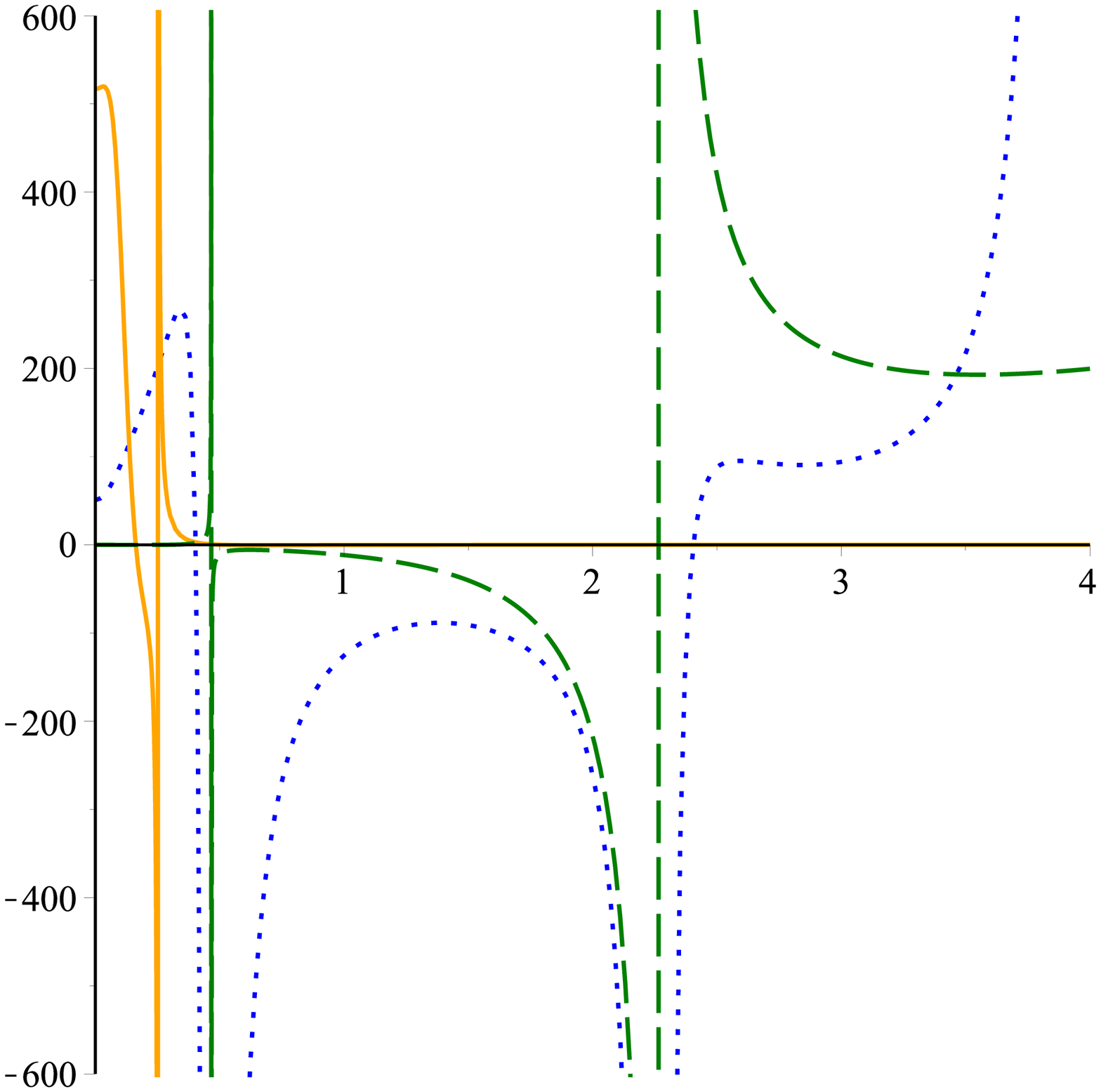}
    }
    \subfigure[]{
        \includegraphics[width=0.4\textwidth]{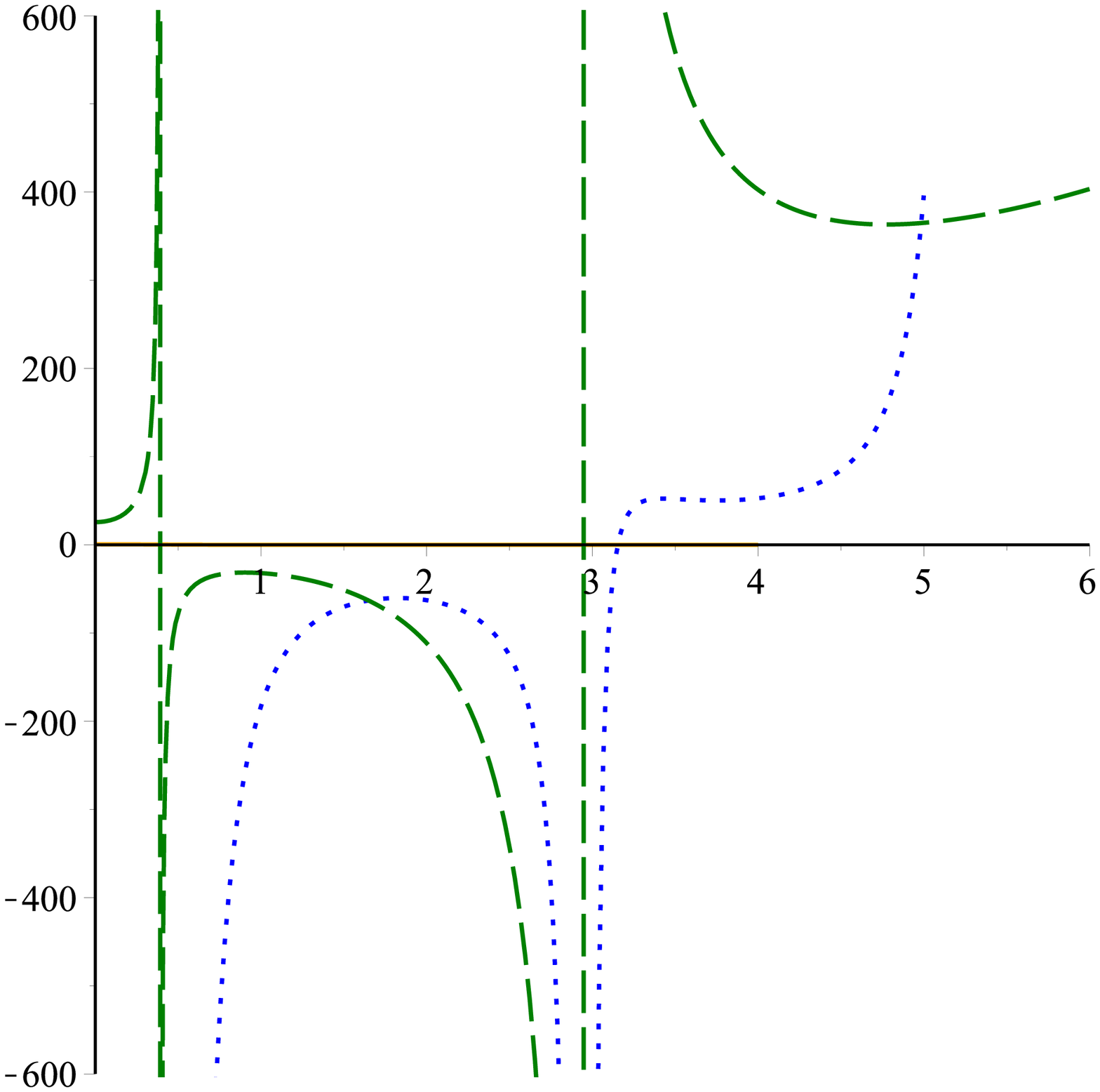}
    }
    \subfigure[]{
        \includegraphics[width=0.4\textwidth]{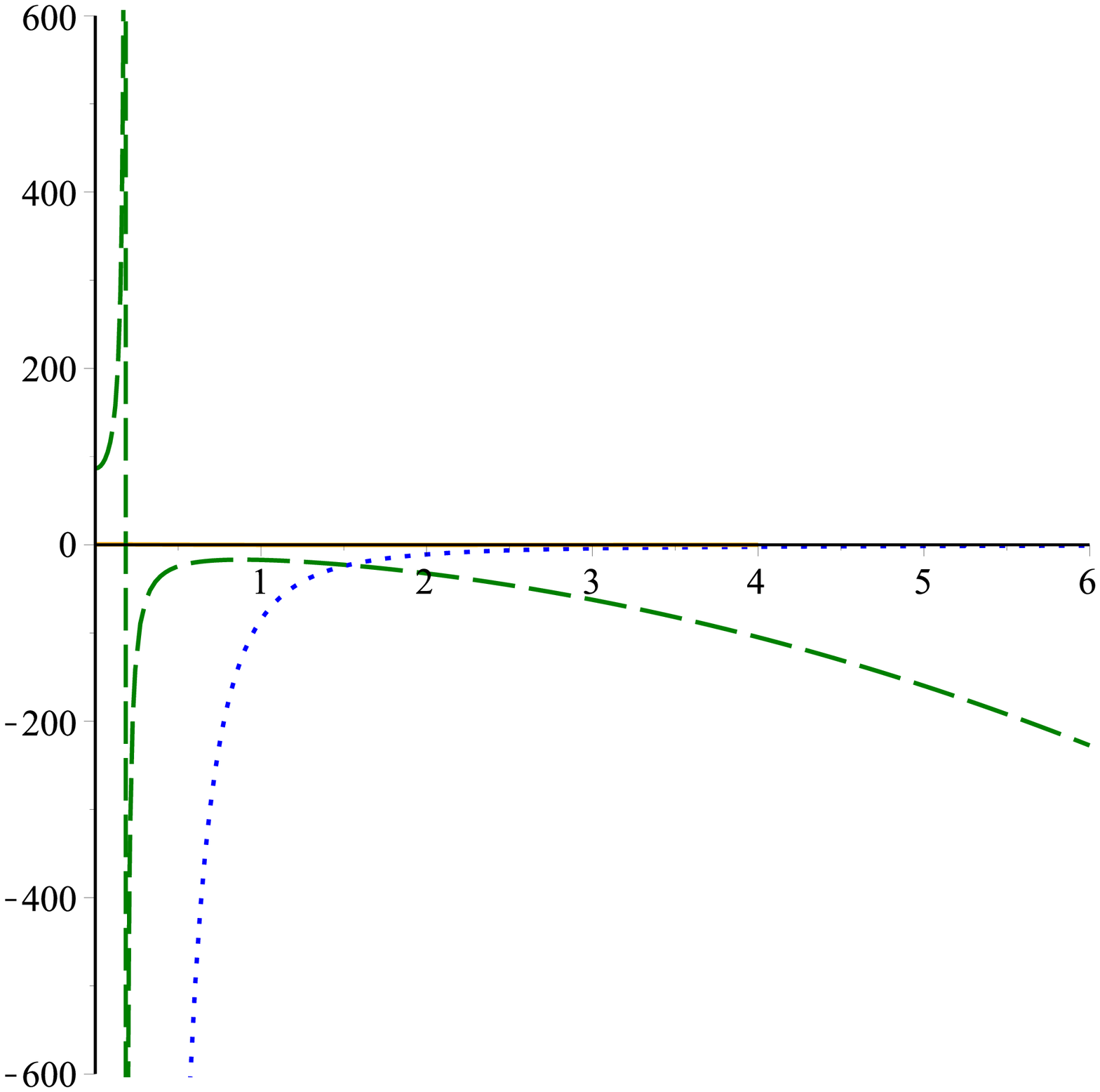}
    }
    \subfigure[]{
        \includegraphics[width=0.4\textwidth]{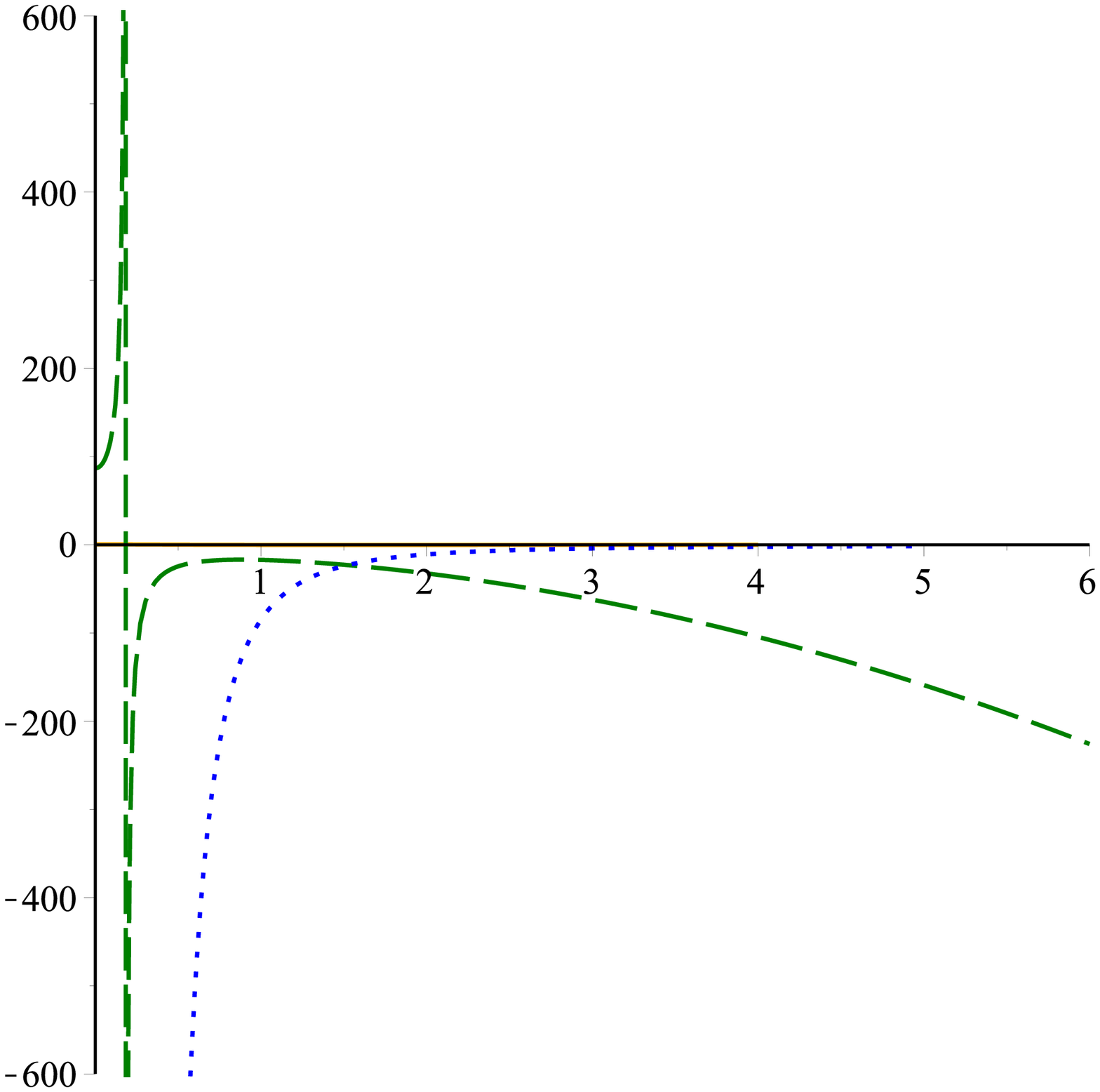}
    }
    \caption{Curvature scalar variation of Ruppeiner (orange continuous line), GTD (blue dot line) metrics, and the heat capacity (green dash line) of a rotating black hole in terms of $ r_{+} $ for $ q=0.25 $, $ a=0.1 $ and $ l=4.0 $, $ l=\sqrt{30} $, $ l=\sqrt{3\cdot 10^{4}} $, $ l=\sqrt{3\cdot 10^{15}} $, for (a), (b), (c) and (d), respectively.}
 \label{pic:CRl}
 \end{figure}

In Figs.~\ref{pic:CRq}--\ref{pic:CRl}, we plot curvature scalar of
Ruppeiner and GTD metrics with the heat capacity of this black hole.
It can be seen from Figs.~\ref{pic:CRq}(a),~\ref{pic:CRa}(a)
and~\ref{pic:CRl}(a), that, this thermodynamical system has one
phase transition type one and two phase transition type two. The
number of these phase transitions changes for different value of 
$ q $, $ a $ and 
$ l $, parameters. By increasing the value of $ q $, the
number of phase transitions will be decreased, as it can be observed
from Fig.~\ref{pic:CRq}(c,d), it has only one phase transition type
one. Also, by increasing the value of $ a $, the number of phase
transitions will be decreased, as it shown in Fig.~\ref{pic:CRa}(d),
it has two phase transition type two. Moreover,  By increasing the
value of $ l $, the number of phase transitions will be decreased,
and it  has only one phase transition type two (see
Fig.~\ref{pic:CRl}(c,d)).

  \clearpage

\section{conclusion}\label{section5}

In this paper, we studied thermodynamic behavior of three types
(static, charged static and charged rotating) of black holes in $
f(R) $ gravity, and investigated the thermodynamic geometry of them.
Also, we plotted thermodynamic quantities in terms of horizon radius
$ r_{+} $ and, we showed that for each maximum and minimum value of
mass, these black holes have one zero point in their temperature and
heat capacity. When we applied the thermodynamic geometry methods to
these black holes, we have seen that, for static black hole,
Weinhold metric is flat, and Ruppiener metric can explain the zero
points of it. For the static charged black hole, Weinhold and
Ruppeiner metrics coincide with the zero points of heat capacity,
and GTD metric can explain the divergence point of it, as well.
Moreover, for the rotating charged black hole, Weinhold metric has no
singularity, but, Ruppiener metric can explain the zero points of
heat capacity and GTD metric coincides with divergence point of it.

We also, investigated the effects of different values of spacetime parameters on
stability conditions of these black holes. We observed that, by changing in
value of spacetime parameters, the number of phase transitions of
these black holes is changed. But these changes has not affected on
compatibility of explained thermodynamical geometry methods with
zeros and divergence points of heat capacity.

For future work, it would be interesting to apply these methods to
other spacetimes such as dilaton black holes.

\bibliographystyle{amsplain}

\end{document}